\documentclass[a4paper,fleqn]{cas-dc}
\usepackage{subcaption}
\usepackage[authoryear,round]{natbib}
\usepackage{amsmath,amssymb} 
\usepackage{graphicx} 
\usepackage{booktabs} 
\usepackage{multirow} 
\usepackage{algorithm} 
\usepackage{algpseudocode} 
\usepackage{xcolor} 
\usepackage{hyperref}
\usepackage{xurl} 
\usepackage{lineno}
\usepackage{pgfplots}
\usepackage{tikz}
\usepackage{xcolor}
\usepackage{mathtools}
\usepackage[most]{tcolorbox}
\usepackage{listings}
\usepackage{array}
\usepackage{rotating}
\usepackage{tabularx}
\usepackage[T1]{fontenc}
\usepackage{amsmath}
\usetikzlibrary{arrows.meta,calc,shapes.geometric}

\lstdefinestyle{promptstyle}{
    basicstyle=\ttfamily\scriptsize,
    breaklines=true,
    columns=fullflexible,
    keepspaces=true,
    showstringspaces=false
}

\newtcolorbox{promptbox}[1]{
    enhanced,
    breakable,
    colback=gray!3,
    colframe=gray!55,
    title=#1,
    fonttitle=\bfseries,
    boxrule=0.6pt,
    arc=2mm,
    left=1mm,
    right=1mm,
    top=1mm,
    bottom=1mm
}

\newtcolorbox{feedbackbox}[1]{
    enhanced,
    breakable,
    colback=blue!2,
    colframe=blue!45!black,
    title=#1,
    fonttitle=\bfseries,
    boxrule=0.6pt,
    arc=2mm,
    left=1mm,
    right=1mm,
    top=1mm,
    bottom=1mm
}

\newtcolorbox{examplebox}[1]{
    enhanced,
    breakable,
    colback=green!2,
    colframe=green!45!black,
    title=#1,
    fonttitle=\bfseries,
    boxrule=0.6pt,
    arc=2mm,
    left=1mm,
    right=1mm,
    top=1mm,
    bottom=1mm
}

\newcommand{\id}[1]{\texttt{#1}}
\newcommand{\pos}[1]{\left\langle #1 \right\rangle}

\renewcommand{\arraystretch}{1.15}

\begin{document}

\shorttitle{OperAItor}

\title[mode=title]{OperAItor: An Agentic Framework for Fault-Tolerant Operation of Chemical Processes}

\author[1]{Javal Vyas}[orcid=0009-0009-6800-8846]
\cormark[1]
\ead{j.vyas24@imperial.ac.uk}

\credit{Conceptualization, Methodology, Software, Investigation, Validation, Writing -- original draft}

\author[1]{Mehmet Mercang{\"o}z}[orcid=0000-0002-4449-0414]
\ead{m.mercangoz@imperial.ac.uk}

\credit{Supervision, Funding Aquistion, Writing -- review and editing}

\affiliation[1]{
  organization={Autonomous Industrial Systems Laboratory, Department of Chemical Engineering,
  Imperial College London},
  addressline={Imperial College Road, South Kensington},
  city={London},
  postcode={SW7 2AZ},
  country={United Kingdom}
}

\cortext[cor1]{Corresponding author}

\begin{abstract}
This paper presents OperAItor, an open-source framework for developing
and evaluating large language model (LLM) agents for supervisory fault
management in chemical processes. The framework is demonstrated on a
dynamic reactor--distillation process representing isopropanol
dehydrogenation, with physically based thermodynamics and modified
reaction kinetics to permit liquid-phase operation. The process is
represented by a 113-variable differential--algebraic model and
operates at a nominal feed rate of 100~kmol/h, producing approximately
4~t/h of acetone. The simulated plant includes eight PI control loops,
22 process alarms, five duty--standby pump trains, and 11 independent
safety instrumented functions. The agent architecture integrates
multivariate statistical process monitoring, fault diagnosis,
knowledge-graph retrieval, action planning, and supervisory action
generation. Five action-validation configurations are investigated,
comprising operation without validation, an LLM-based judge,
digital-twin simulation, and deterministic admissibility checks with
and without consequence projection. The framework is evaluated using
four LLMs across 20 scenarios involving single and multiple simultaneous
faults. Without agent intervention, 18 scenarios result in plant stops,
corresponding to an economic loss of \$399,000 relative to fault-free
operation under the assumed benchmark economics. For GPT-5 nano,
Qwen3 8B, and Mistral 7B, validated agent configurations reduce the
number of stops to 5--8 and recover \$209,000--265,000 of the loss.
For Llama 3.2 3B, deterministic admissibility checks reduce the number
of stops from 13 without validation to 7, whereas LLM-based validation
results in 15 stops. An offline admissibility analysis finds no feasible recovery among
the evaluated single supervisory actions for four condenser-cooling-loss
scenarios. The study provides a reproducible
benchmark for comparing agent architectures, language models, and
action-validation methods for supervisory fault management in
chemical processes.
\end{abstract}

\begin{highlights}
\item Hypothetical reactor--distillation benchmark with process control, alarms, and 11 independent SIFs.
\item Five validation configurations are compared for LLM-generated supervisory actions.
\item Twenty single and multiple-fault scenarios are assessed using plant stops and illustrative economics.
\item Deterministic admissibility checks improve outcomes for the 3B-parameter language model.
\item Offline admissibility analysis identifies four condenser-loss cases with no feasible single-action recovery.
\end{highlights}


\maketitle


\section{Introduction}
\label{sec:intro}

Abnormal operating conditions in chemical processes often require
interventions that extend beyond the capabilities of regulatory
control. A failed duty pump may require the activation of standby
equipment, a deteriorating heat exchanger may require a reduction
in process throughput, and an actuator fault may require an
alternative operating strategy. Although the basic process control
system (BPCS) regulates normal operation and an independent safety
instrumented system (SIS) provides protection against specified
hazardous conditions, neither necessarily determines the most
appropriate corrective intervention following a fault
\citep{isa182,iec61511}. This responsibility typically rests with
plant operators, who must interpret alarms, identify the likely
cause of the abnormal condition, assess available interventions,
and act before the process reaches a protective shutdown.
The challenge is to detect abnormal operation and select
corrective actions that sustain production while respecting
process constraints.

Fault detection and diagnosis (FDD) and fault-tolerant control
(FTC) have addressed different aspects of this problem.
Statistical process monitoring, including principal component
analysis (PCA), provides established methods for detecting
deviations from nominal operation
\citep{macgregor1995,qin2003,venkat2003c,chiang2001}.
Model-based diagnosis and active FTC extend this capability
towards fault isolation and control reconfiguration
\citep{venkat2003a,blanke2006,zhang2008ftc,mhaskar2013}.
However, the availability of a fault diagnosis does not
automatically determine an appropriate operator intervention.
Recovery may require discrete equipment operations, changes
to supervisory setpoints, or coordinated adjustments across
interacting process units. The feasibility of an intervention
depends on the process state, actuator availability, operating
procedures, intervention authority, and the consequences of the
proposed action.
These considerations motivate supervisory decision-making
frameworks that connect process monitoring and diagnosis
with executable, plant-specific recovery actions.

Large language models (LLMs) offer a potential mechanism for
such supervisory decision-making. Through structured interfaces
and access to plant-specific information, LLM-based agents
can interpret alarms and operating trends, retrieve equipment
relationships and procedures, formulate recovery plans, and
translate these plans into operator-level commands
\citep{brown2020,yao2023react,decardinelson2024,schweidtmann2024}.
Previous studies have demonstrated the feasibility of using
LLM agents for industrial process control and fault recovery,
including architectures that combine knowledge retrieval,
planning, and simulation-based action validation
\citep{vyas2024autonomous,vyas2026detection}.
Related work has investigated deterministic verification of
LLM-generated industrial actions using engineering information
such as piping and instrumentation diagrams (P\&IDs) and
operating constraints \citep{schall2026pid}. These developments
establish that LLMs can generate meaningful supervisory
interventions when their access to the process is appropriately
constrained. They do not, however, eliminate the possibility
of incorrect fault attribution, unsuitable control actions,
or interventions that are individually executable but
undesirable when applied to the current process state.

This distinction motivates the central question investigated
in this work: \emph{given an LLM-generated operator action,
how should its suitability for execution be evaluated?}
The question is important because action generation and
action validation address different problems. An agent may
correctly identify a fault yet recommend an intervention that
accelerates a protective trip. Conversely, a validation
mechanism may reject a beneficial intervention because its
acceptance conditions are overly restrictive. A useful
validation layer must therefore balance the exclusion of
undesirable actions against the preservation of opportunities
for fault recovery. Its effectiveness cannot be established
solely by measuring the validity of generated text or the
proportion of proposals accepted. It must also be assessed
through the resulting closed-loop process behaviour.

Several validation approaches can be considered. A
\emph{model-based validator} evaluates a proposed action by
simulating its predicted consequences using a process model
or digital twin, allowing candidate interventions to be
compared before execution \citep{kritzinger2018,wabersich2021}.
An \emph{admissibility-based validator} evaluates whether
the proposed intervention is supported by appropriate
evidence, permitted within the agent's authority, and
consistent with operating procedures and predefined
constraints. The ADMIT framework provides a structured
approach to such action-level evaluation through a sequence
of deterministic checks, optionally supplemented by
consequence projection \citep{misra2026admitbench}.
Alternatively, an \emph{LLM-based judge} can examine the
proposed intervention and associated process information
and produce an acceptance decision without requiring an
explicit process simulator \citep{zheng2023judge}.
These approaches differ in their information requirements,
computational expense, rejection criteria, and ability to
evaluate the consequences of an intervention. Whether these
differences translate into improved fault recovery and
economic performance remains an important empirical question.

A controlled comparison requires a process benchmark in
which candidate interventions have observable consequences
for both process operation and protective systems.
Established benchmarks such as the Tennessee Eastman
process have supported extensive research on fault
detection, process monitoring, and control
\citep{downs1993,reinartz2021tep}. However, conventional
fault-detection benchmarks do not necessarily provide
the complete supervisory environment needed to evaluate
operator-level actions, including equipment redundancy,
explicit action permissions, alarm handling, and
independent protective interlocks. Furthermore, detection
accuracy alone does not capture whether an intervention
prevents a plant stop or preserves production. An action-
validation benchmark must therefore represent process dynamics,
fault mechanisms, the operational interface for executing
decisions, and the criteria used to assess their consequences.

To address these requirements, this paper presents
\emph{OperAItor}, an open-source framework and benchmark
for evaluating LLM-generated supervisory interventions
during abnormal chemical process operation. The case study
uses a hypothetical reactor--distillation process with physically
based thermodynamics and modified liquid-phase reaction kinetics,
represented by a 113-variable differential--algebraic equation
(DAE) model. It incorporates eight
regulatory PI control loops, 22 process alarms,
five duty/standby pump trains, and eleven simulated
safety instrumented functions. The agentic architecture
connects PCA-based monitoring, fault diagnosis,
knowledge-graph retrieval, planning, and LLM-generated
operator actions through a modular validation interface.
The agents operate above the BPCS and are not permitted
to bypass the independent protective logic.

The framework is used to compare five action-validation
configurations: no validation beyond command executability,
an LLM judge, a digital-twin validator, deterministic
ADMIT admissibility gates, and ADMIT gates supplemented
by consequence projection. These configurations are
evaluated using four LLMs across twenty scenarios
containing individual, coupled, and triple faults.
All configurations use the same simulated plant,
supervisory action interface, fault scenarios, and
benchmark scoring procedure. Performance is evaluated
using plant-stop frequency, operational economic
performance, action rejection behaviour, and
computational cost. An additional offline admissibility
analysis examines the recovery limitations imposed
by the available supervisory actions. The digital-twin
experiments use a matched-model setting, allowing the
comparison to isolate validation behaviour without
introducing plant--model mismatch.

The principal contributions of this work are:

\begin{enumerate}

\item \textbf{An integrated benchmark for agentic
fault-tolerant process operation.}
We develop an open dynamic reactor--distillation
benchmark that combines process dynamics, regulatory
control, alarm management, equipment redundancy,
and simulated safety interlocks. A common operator
action interface enables alternative supervisory
decision-making methods to be evaluated under
identical process conditions
(Sections~\ref{sec:case}--\ref{sec:control}).

\item \textbf{A modular framework for generating
and validating supervisory interventions.}
OperAItor integrates statistical monitoring,
fault diagnosis, process knowledge retrieval,
planning, and LLM-based action generation with
interchangeable validation mechanisms. The framework
separates action proposal, validation, and execution,
allowing their individual roles to be examined
without changing the underlying plant model
(Section~\ref{sec:framework}).

\item \textbf{A comparative evaluation of
action-validation strategies.}
Five validation configurations are assessed
using four LLMs and twenty fault scenarios,
including interacting faults. The evaluation
quantifies how alternative validation mechanisms
affect fault recovery, plant stops, operational
economic performance, rejected interventions,
and computational requirements
(Sections~\ref{sec:benchmark}--\ref{sec:results}).

\item \textbf{An analysis of validation and
recovery limitations.}
By examining rejection outcomes and offline
admissibility results, we distinguish limitations
associated with the action-generation and validation
mechanisms from faults that cannot be resolved
through the available individual supervisory
interventions. This analysis provides a basis
for interpreting benchmark performance beyond
action acceptance rates alone
(Section~\ref{sec:results}).

\end{enumerate}

LLM-based agents offer opportunities to automate supervisory
decisions in chemical process operations, particularly in
responding to abnormal operating conditions and process faults.
However, actions proposed by these agents may be inappropriate
for the current plant state, violate operating constraints, or
lead to undesirable process behaviour. Methods for validating
proposed actions before their implementation are therefore an
important component of agentic supervisory systems. Evaluating
these methods requires a dynamic process environment in which
the consequences of supervisory interventions can be assessed
under repeatable fault conditions. This study introduces
OperAItor, an open-source framework that combines a dynamic
process simulator, fault scenarios, LLM-based supervisory
agents, and interchangeable action-validation mechanisms.
The framework enables systematic comparisons of agent and
validation configurations in terms of operational performance,
fault recovery, and plant stops. The simulator, agent framework,
fault scenarios, and evaluation records are made publicly
available to support reproducible research \citep{operaitor_github,operaitor_site}.
\section{Case study: An integrated reactor--distillation benchmark}
\label{sec:case}

\begin{figure*}[t]
\centering
\includegraphics[width=\textwidth]{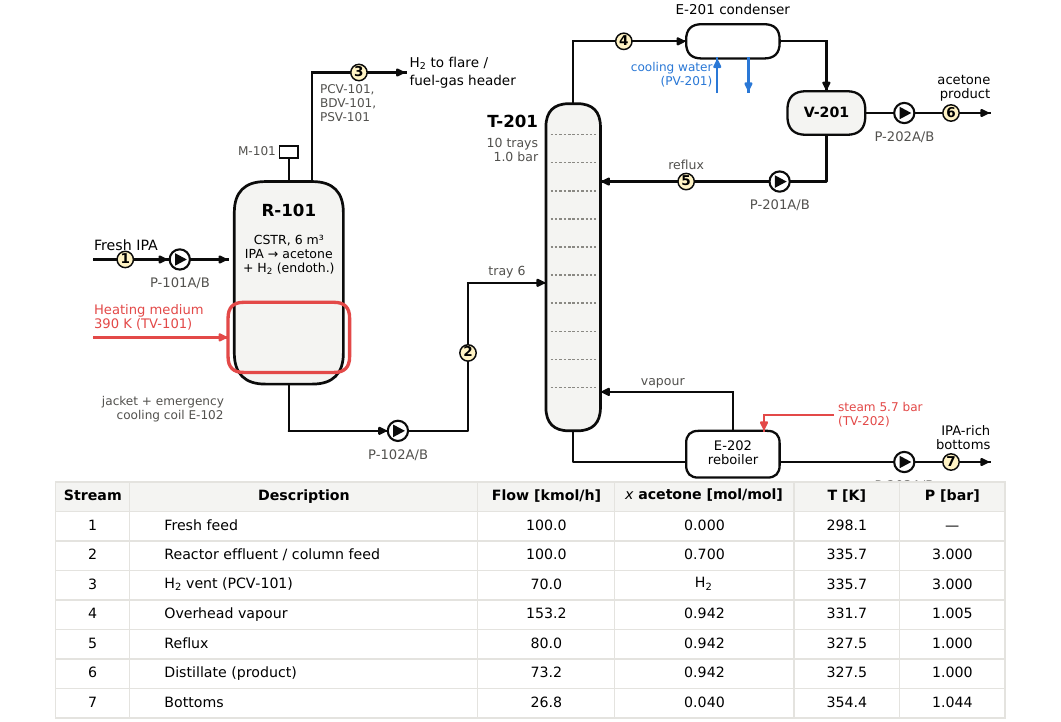}
\caption{Process flow diagram and nominal stream data for the reactor--distillation benchmark. Five liquid services use duty/standby pump trains. Detailed instrumentation and protective interlocks are presented in Section~\ref{sec:control}.}
\label{fig:pfd}
\end{figure*}

Evaluating an LLM-generated supervisory intervention requires a process in which the intervention has observable consequences for material inventories, operating constraints, and protective systems. We therefore consider the integrated reactor--distillation process in Fig.~\ref{fig:pfd}. The benchmark combines continuously evolving process states with discrete equipment failures, redundant pumps, regulatory control, and protective interventions. It is designed to assess fault recovery through operator-level actions rather than to reproduce the detailed design of a particular industrial facility.

The process represents the generic transformation $A\rightarrow B+C$, with the gaseous product $C$ removed from the reactor and the liquid $A$--$B$ mixture separated downstream. The species used in the simulation are isopropanol (IPA, $A$), acetone ($B$), and hydrogen ($C$), following the stoichiometry of IPA dehydrogenation \citep{turton2012,luyben2011acetone},
\begin{equation}
\mathrm{(CH_3)_2CHOH}\longrightarrow\mathrm{(CH_3)_2CO}+\mathrm{H_2},
\qquad \Delta H_r=+68\ \mathrm{kJ\,mol^{-1}}.
\label{eq:reaction}
\end{equation}
Fresh IPA enters the jacketed continuous stirred-tank reactor R-101 through the duty/standby feed pumps P-101A/B. At the nominal operating point, the fresh feed is 100~kmol\,h$^{-1}$, the reactor temperature is 335.74~K, the pressure is 3.0~bar, and the conversion is 70\%. The heating jacket supplies the endothermic reaction duty, while a separate emergency-cooling circuit can be actuated by the protective system. Hydrogen is discharged through the pressure-control valve PCV-101; an independently actuated blowdown valve and a pressure-relief-valve surrogate provide additional discharge paths.

The reactor effluent is transferred by P-102A/B to tray~6 of the ten-tray distillation column T-201. Overhead vapour is condensed in E-201 and collected in reflux drum V-201. The column has separate duty/standby trains for reflux, distillate withdrawal, and bottoms withdrawal (P-201A/B, P-202A/B, and P-203A/B, respectively). Cooling-water flow to the condenser and steam flow to the kettle reboiler E-202 provide the principal thermal manipulations. The nominal column pressure is 1.0~bar, with a reflux rate of 80.0~kmol\,h$^{-1}$, a distillate rate of 73.2~kmol\,h$^{-1}$, and a bottoms rate of 26.8~kmol\,h$^{-1}$. Their acetone mole fractions are 0.942 and 0.040, respectively. The nominal fresh-feed rate corresponds to approximately 6.01~t/h of IPA, while 70\% reactor conversion gives approximately 4.07~t/h of acetone formation. At the stated distillate rate and composition, the acetone product rate is approximately 4.0~t/h. The nominal stream calculations report an acetone-balance closure within $10^{-2}$~kmol\,h$^{-1}$ when evaluated using the unrounded simulated stream values.

\paragraph{Physical interpretation.}
The process uses thermophysical properties of the specified chemical
species, while the dehydrogenation rate is modified to permit reaction
in the liquid phase at approximately 336~K. Industrial IPA
dehydrogenation normally takes place in the gas phase at substantially
higher temperatures \citep{turton2012,luyben2011acetone}. The
reaction kinetics therefore serve as a dynamic surrogate for a
hypothetical endothermic $A\rightarrow B+C$ process. The model is used
to evaluate causal process responses, operator-level interventions, and
protective functions rather than to predict the operation of a
commercial acetone plant.

\subsection{Dynamic process model}

The plant is represented by coupled component, total-holdup, and energy balances, with algebraic constraints for thermodynamics, hydraulics, and equipment behaviour. Acetone is the light component and IPA the heavy component in the separation model. Vapour pressures are represented by a two-parameter Antoine correlation, and non-ideal liquid-phase behaviour by an asymmetric Margules activity-coefficient model \citep{poling2001}. The vapour--liquid equilibrium closure is
\begin{equation}
P_{\mathrm{bub}}(T,x)=\sum_{j=1}^{2}x_j\gamma_j(x)P_j^{\mathrm{sat}}(T),
\qquad
y_1=\frac{x_1\gamma_1(x)P_1^{\mathrm{sat}}(T)}{P},
\end{equation}
where $x_j$ and $y_j$ are liquid and vapour mole fractions. Liquid and vapour enthalpies use constant heat capacities and component heats of vaporisation. The complete property correlations and numerical parameter values are retained in the released implementation.

\paragraph{Reactor dynamics.}
Reactor R-101 has four differential states: total liquid holdup $M$, acetone holdup $n_B$, internal energy $U_R$, and gas-phase hydrogen holdup $n_{H_2}$. Reaction rate $r$ depends on the unconverted IPA inventory and an Arrhenius rate coefficient,
\begin{equation}
r=a\,k_{\mathrm{ref}}
\exp\!\left[-\frac{E_a}{R_g}
\left(\frac{1}{T_R}-\frac{1}{T_{\mathrm{ref},k}}\right)\right]
\pos{M-n_B},
\end{equation}
where $a=1$ at nominal catalyst activity and $\pos{z}=\tfrac12(z+\sqrt{z^2+\varepsilon^2})$ is a smooth positive-part approximation with $\varepsilon=10^{-10}$. The principal balances are
\begin{align}
\dot M &= F-(F_{\mathrm{out}}+F_{\mathrm{leak}}),
\label{eq:cstr_M}\\
\dot n_B &= r-(F_{\mathrm{out}}+F_{\mathrm{leak}})x_R,\\
\dot n_{H_2} &= r-F_{\mathrm{vent}},\\
\dot U_R &= Fh^L(T_F,0)
 -(F_{\mathrm{out}}+F_{\mathrm{leak}})h^L(T_R,x_R)
 \notag\\
&\quad -F_{\mathrm{vent}}h_{H_2}(T_R)
 +Q_j-Q_{\mathrm{loss}}-Q_{\mathrm{em}}
 -r\Delta H_r.
\label{eq:cstr_U}
\end{align}
Here $F$ is the feed rate, $F_{\mathrm{out}}$ is the liquid effluent, and $F_{\mathrm{leak}}$ is a fault-induced leak. The jacket, ambient heat loss, and emergency cooling contribute $Q_j$, $Q_{\mathrm{loss}}$, and $Q_{\mathrm{em}}$. The algebraic relations include $x_R=n_B/M$ and $P_R=n_{H_2}R_gT_R/V_g$, with an effective gas capacitance represented through $V_g$. Hydrogen discharge depends on the actual openings of the pressure-control and blowdown valves and on a simplified relief-valve characteristic. These relations allow reduced vent capacity, pressure-sensor bias, and a heater stuck at high output to produce distinct dynamic responses.

\paragraph{Distillation and inventory dynamics.}
Column T-201 is a ten-tray binary equilibrium-stage model with a kettle reboiler, a reflux drum, and a lumped condenser-pressure state, following conventional dynamic column formulations \citep{luyben1990,skogestad1997}. Each tray $i$ contains liquid holdup $M_i$, acetone holdup $n_i$, and internal energy $U_i$. With liquid flowing downward and vapour upward, its material balances are
\begin{align}
\dot M_i &= L_{i-1}+V_{i+1}+F_i-L_i-V_i,\\
\dot n_i &= L_{i-1}x_{i-1}+V_{i+1}y_{i+1}
          +F_i z-L_i x_i-V_i y_i,
\end{align}
where $F_i$ is non-zero only on feed tray~6. Tray energy balances account for the corresponding liquid, vapour, and feed enthalpy flows. Tray temperatures and phase compositions are determined algebraically from the vapour--liquid equilibrium relations, while liquid and vapour flows follow holdup- and pressure-dependent hydraulic correlations. The reboiler includes steam heating and bottoms withdrawal.

The condenser model includes finite cooling capacity. With cooling-water valve position $u_c$, effectiveness factor $\eta$, and top-tray temperature $T_1$, its available duty, condensation rate, and pressure dynamics are
\begin{align}
Q_{\mathrm{avail}} &= UA_c\,u_c\,\eta\pos{T_1-T_{cw}},\\
F_{\mathrm{cond}} &= \min\!\left(
\frac{Q_{\mathrm{avail}}}{\lambda_c},
\pos{V_1+k_P(P_c-1.0)}\right),\\
\dot P_c &= \frac{V_1-F_{\mathrm{cond}}}{C_P}.
\label{eq:condenser}
\end{align}
Here $P_c$ is column-overhead pressure, $\lambda_c$ is condensation enthalpy, and $C_P$ is an effective vapour capacitance. A reduction in $\eta$ limits condensation, raises $P_c$, and can initiate a protective high-pressure trip. The reflux-drum liquid inventory follows
\begin{equation}
\dot M_d=F_{\mathrm{cond}}-(R+D+F_{\mathrm{leak}}^d),
\end{equation}
where $R$ and $D$ denote reflux and distillate withdrawal. This balance makes pump trips and drum leakage directly relevant to the feasibility of supervisory recovery actions.

\paragraph{Coupling and numerical solution.}
The reactor liquid outflow supplies the column feed, giving a single semi-explicit, index-1 differential--algebraic system,
\begin{equation}
\mathbf F\!\left(t,\mathbf y,
\dot{\mathbf y}_{\mathcal D},\mathbf u\right)=\mathbf 0,
\qquad \mathbf y\in\mathbb R^{113},
\label{eq:dae}
\end{equation}
with 41 differential and 72 algebraic variables. The differential states comprise four reactor states, three states on each of ten trays, three states each in the reboiler and reflux drum, and one condenser-pressure state. The system is integrated by implicit backward Euler at a nominal process step of 0.1~min. Each step is solved using Powell's hybrid method with scaled residuals and adaptive step subdivision when convergence is difficult \citep{powell1970,virtanen2020scipy}. A scaled residual norm below $10^{-5}$ is required; if repeated subdivision and the final least-squares fallback fail, the episode is terminated as a numerical failure. Such terminations are counted as plant stops in the benchmark and are therefore distinguished from physical protective trips in the recorded outcomes. Regulatory controllers update every 1~min, while process observations used by the monitoring and evaluation layers are sampled every 0.5~min. Numerical details and the full parameter set are available with the simulator.

\subsection{Fault mechanisms and disturbances}
\label{sec:faults}

A central feature of the benchmark is the separation between a requested intervention and the action actually delivered to the process. For each manipulated variable, the simulator distinguishes the BPCS request, the supervisory request, the safety-arbitrated command, and the final plant input. Actuator and equipment faults affect this last stage. As a result, a safety interlock may demand a prescribed valve movement without being able to restore the failed final element. The arbitration rules and independent protective logic are described in Section~\ref{sec:control}.

The fault library contains 18 named fault mechanisms covering actuator, sensor, equipment, and process faults. Ten are used in the twenty-scenario evaluation campaign (Table~\ref{tab:faults}). Equipment trips remove flow from the duty pump until the standby pump is started. Actuator faults impose capacity restrictions or fixed positions; sensor faults bias the measurements available to regulatory control; and process faults modify reaction activity, heat-transfer effectiveness, or material inventories. These interventions enter the differential and algebraic equations directly rather than being represented only as alarm labels.

\begin{table*}[t]
\centering
\caption{Faults selected for the benchmark campaign. Machine-readable fault identifiers are provided with the released scenarios.}
\label{tab:faults}
\small
\begin{tabularx}{\textwidth}{@{}p{3.7cm}p{1.9cm}p{1.8cm}X@{}}
\toprule
Fault & Class & Element & Injected mechanism \\
\midrule
Feed pump trip & Equipment & P-101A & Duty pump stops; P-101B remains available. \\
Reflux pump trip & Equipment & P-201A & Duty pump stops; P-201B remains available. \\
Distillate pump trip & Equipment & P-202A & Duty pump stops; P-202B remains available. \\
Bottoms pump trip & Equipment & P-203A & Duty pump stops; P-203B remains available. \\
Reactor outlet capacity loss & Actuator & FV-102 & Maximum achievable flow reduced to 20\% of design. \\
Hydrogen vent degradation & Actuator & PCV-101 & Achievable opening falls to 20\% over 10~min. \\
Reactor heater stuck high & Actuator & TV-101 & Heating factor fixed at 1.0 (nominal 0.61). \\
Reactor pressure bias & Sensor & PT-101 & BPCS pressure measurement biased by $-1.2$~bar. \\
Condenser cooling loss & Process & E-201 & Cooling effectiveness falls to $\eta=0.2$. \\
Reflux drum leak & Process & V-201 & Liquid inventory loss of 100~kmol\,h$^{-1}$. \\
\bottomrule
\end{tabularx}
\end{table*}

To distinguish injected faults from normal operating variability, the benchmark also permits finite-duration disturbances: 15-min steps of $\pm3$~K in cooling-water and ambient temperature, $\pm5$~K in fresh-feed and heating-medium temperature, and $\pm10$\% in steam-header pressure, together with Gaussian measurement noise. The twenty benchmark scenarios define the combinations, onset times, and fault durations used for comparisons; these are specified in Section~\ref{sec:benchmark}. The following section describes the regulatory and protective layers through which these physical faults become alarms, control responses, and potential plant stops.

\section{Control and protection architecture}
\label{sec:control}

The reactor--distillation benchmark incorporates three operational layers with distinct responsibilities: a basic process control system (BPCS) for regulatory control, an alarm system for detecting and communicating abnormal conditions, and a simulated safety instrumented system (SIS) for protective intervention. The LLM-based operator framework introduced in Section~\ref{sec:framework} operates above these layers. It may modify permitted supervisory targets and initiate equipment operations, but it does not replace regulatory controllers or override the protective logic. This separation is essential for evaluating whether an agent-generated intervention is executable, operationally beneficial, and consistent with the constraints imposed by the plant.

\subsection{Regulatory control}

The BPCS comprises eight single-loop proportional--integral (PI) controllers updated every $\Delta t_c=1$~min. The control structure follows conventional reactor and distillation control principles \citep{skogestad1997,luyben2011acetone}. Reactor liquid inventory is regulated through its outlet flow, reactor temperature through jacket heating, and reactor pressure through hydrogen venting. For the distillation column, top-tray temperature is controlled through reflux, bottom temperature through reboiler steam, and column pressure through condenser cooling water. Reflux-drum and reboiler inventories are regulated by distillate and bottoms withdrawal, respectively. Fresh-feed throughput is specified separately through FIC-101 as a supervisory production target.

Each controller implements a positional PI law with a measurement deadband. Denoting the setpoint and measured process variable at control interval $k$ by $SP_k$ and $PV_k$, the control error and output are
\begin{align}
e_k &=
\begin{cases}
0, & |SP_k-PV_k|\leq\delta,\\
SP_k-PV_k, & \text{otherwise},
\end{cases}
\\
u_k^{\mathrm{BPCS}}
&=
\operatorname{clip}_{[u_{\min},u_{\max}]}
\left(b+K_pe_k+I_k\right),
\label{eq:bpcs_pi}\\
I_{k+1}
&=
I_k+\alpha_k\frac{K_p}{T_i}e_k\Delta t_c,
\end{align}
where $K_p$ is the proportional gain, $T_i$ is the integral time, $b$ is the nominal controller-output bias, and $\delta$ is the deadband. The conditional integration indicator $\alpha_k\in\{0,1\}$ prevents further integral accumulation when the output is saturated and the proposed integral update would increase saturation. Bumpless transfer from manual to automatic operation is implemented by initializing the integral state as $I=u_{\mathrm{manual}}-b-K_pe$ \citep{astrom2006}.

The controller pairings, tuning parameters, permissible supervisory setpoint ranges, and closed-loop response characteristics are summarised in Table~\ref{tab:loops}. The setpoint limits define the range within which an operator or agent may request a change; they do not guarantee that the resulting process trajectory will remain within safe operating limits.

\begin{table*}[t]
\centering
\caption{Regulatory control configuration and closed-loop step-test performance. OS denotes percentage overshoot and $t_s$ the settling time within 20\% of the imposed step, or two deadbands.}
\label{tab:loops}
\small
\setlength{\tabcolsep}{4pt}
\begin{tabularx}{\textwidth}{@{}lXrrllrr@{}}
\toprule
Loop & Controlled variable $\rightarrow$ manipulated variable
& $K_p$ & $T_i$ [h] & Deadband & Setpoint range
& OS [\%] & $t_s$ [min]\\
\midrule
\id{R\_LIC} &
Reactor holdup $\rightarrow$ outlet flow &
$-6.0$ & 2.00 & 0.50 kmol & 40--60 kmol & 4 & 9.5\\
\id{R\_TIC} &
Reactor temperature $\rightarrow$ heating &
$+0.030$ & 0.12 & 0.30 K & 325--340 K & 6 & 10.5\\
\id{R\_PIC} &
Reactor pressure $\rightarrow$ H$_2$ vent &
$-0.10$ & 0.30 & 0.015 bar & 2.5--3.4 bar & 0 & 33.0\\
\id{TOP\_TIC} &
Top-tray temperature $\rightarrow$ reflux &
$-7.0$ & 0.15 & 0.30 K & 325--340 K & 8 & 10.5\\
\id{COL\_PIC} &
Column pressure $\rightarrow$ cooling water &
$-4.5$ & 0.20 & 0.002 bar & 0.90--1.12 bar & 20 & 8.5\\
\id{DR\_LIC} &
Drum holdup $\rightarrow$ distillate &
$-12.0$ & 2.00 & 0.30 kmol & 16--40 kmol & 8 & 8.0\\
\id{BOT\_TIC} &
Bottom temperature $\rightarrow$ steam &
$+0.0325$ & 0.15 & 0.30 K & 345--365 K & 0 & 15.5\\
\id{RB\_LIC} &
Reboiler holdup $\rightarrow$ bottoms &
$-6.0$ & 2.00 & 0.35 kmol & 18--46 kmol & 7 & 14.0\\
\bottomrule
\end{tabularx}
\end{table*}

\begin{figure*}[pos=htbp]
\centering
\includegraphics[width=\textwidth]{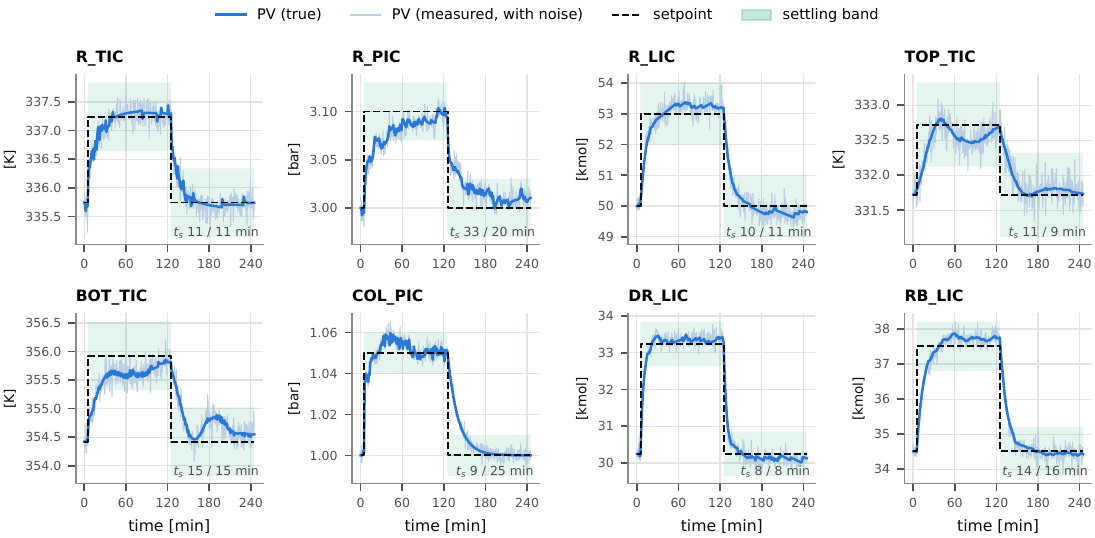}
\caption{Closed-loop setpoint responses of the eight regulatory PI controllers. Setpoint changes are introduced at 5~min and reversed at 125~min. The traces include noisy BPCS measurements.}
\label{fig:loops}
\end{figure*}

The regulatory loops were evaluated from the nominal steady state using setpoint changes at 5~min followed by a return to the original setpoint at 125~min, with measurement noise included. Figure~\ref{fig:loops} shows the resulting responses. All eight loops initially move in the intended control direction, settle within the specified tolerance without activating a protective interlock or priority-1 alarm, and return to the nominal operating point. Reported overshoots range from 0 to 20\%, with settling times between 8 and 33~min. The reactor-pressure loop is the slowest, reflecting the effective hydrogen-gas capacitance represented in the model.

Disturbance-rejection tests were also performed for 40-min perturbations of the utility and feed conditions. A $+3$~K cooling-water temperature disturbance produced a maximum column-pressure deviation of 0.011~bar, while the cooling-water valve moved from approximately 60 to 69\% opening. Fresh-feed temperature and steam-header pressure disturbances produced maximum controlled-temperature deviations of 0.16 and 0.35~K, respectively. In each case, the controlled variable subsequently returned to its setpoint. These tests establish the nominal regulatory response against which the fault-recovery interventions are evaluated.

\subsection{Alarm management}

The benchmark includes 22 configured alarms with priorities, activation limits, hysteresis, and acknowledgement states, following the basic alarm-management concepts of ISA-18.2 \citep{isa182}. The alarm catalogue covers abnormal reactor pressure, temperature and liquid inventory; column pressure; reflux-drum, reboiler and tray inventories; product-quality deviations; and pump-train failures. Process alarms are assigned priorities 1--2, product-quality alarms priority 3, and pump-trip alarms priority 1.

Each alarm is activated when its corresponding measured variable crosses the configured threshold. Hysteresis prevents repeated activation and clearing near the limit. For example, a high alarm with threshold $L$ and hysteresis $h$ becomes active at $PV\geq L$ and clears only after $PV<L-h$. The implemented alarm-state sequence distinguishes unacknowledged active alarms, acknowledged active alarms, and return-to-normal conditions that have not yet been acknowledged.

Alarm thresholds are selected to provide a warning region before corresponding protective limits are reached. Consequently, an abnormal condition may first generate an alarm, subsequently activate an interlock, and finally cause a plant stop if the process continues to deteriorate. The alarm system provides information to the operator or LLM agents but does not, by itself, determine a corrective action. An alarm acknowledgement also does not alter the underlying process state or clear a protective interlock.

This distinction matters for supervisory fault recovery. An agent may observe several simultaneous alarms caused by a single equipment fault, while the regulatory controllers may already be saturated in attempting to compensate. The appropriate intervention must therefore account for the underlying process condition rather than treating each alarm as an independent control objective.

\subsection{Protective interlocks and plant-stop logic}

The simulated SIS comprises eleven independently evaluated protective interlocks, denoted I1--I11. Each interlock monitors a specified process variable and, when its activation threshold is reached, imposes predefined overrides on the affected process commands. The interlocks are evaluated using the simulated physical process state independently of the potentially biased or noisy measurements supplied to the BPCS. This separation allows sensor faults to compromise regulatory control without simultaneously corrupting the protective measurements.

The interlocks are latched and require an explicit reset request after the monitored variable has returned inside its reset threshold. The reset hysteresis prevents an interlock from repeatedly activating and clearing close to its limit. Six of the eleven interlocks additionally initiate a latched plant-stop event, which terminates the corresponding benchmark episode. Table~\ref{tab:ce} summarises the activation conditions, principal protective actions, and plant-stop consequences. The complete actuator-level cause-and-effect matrix is retained in the supplementary material and the software release.

\begin{table*}[t]
\centering
\caption{Summary of the eleven protective interlocks. A plant-stop latch terminates the benchmark episode. The full cause-and-effect matrix specifies the actuator-level overrides applied by each interlock.}
\label{tab:ce}
\footnotesize
\setlength{\tabcolsep}{3pt}
\begin{tabularx}{\textwidth}{@{}clllXc@{}}
\toprule
ID & Interlock & Activation & Reset condition &
Principal protective action & Plant stop\\
\midrule
I1 & \id{R\_T\_HH} &
$T_R\geq348$ K & $T_R<345$ K &
Stop reactor heating; activate emergency cooling & No\\

I2 & \id{R\_P\_HH} &
$P_R\geq4.00$ bar & $P_R<3.80$ bar &
Stop heating; fully open hydrogen vent and blowdown & No\\

I3 & \id{R\_P\_HHH} &
$P_R\geq4.35$ bar & $P_R<4.13$ bar &
Limit feed to 40\%; stop heating; open vent and blowdown & Yes\\

I4 & \id{R\_L\_HH} &
$M_R\geq72$ kmol & $M_R<68.4$ kmol &
Stop fresh feed and reactor heating & No\\

I5 & \id{R\_L\_LL} &
$M_R\leq20$ kmol & $M_R>21$ kmol &
Stop reactor heating and reactor liquid outlet & No\\

I6 & \id{RB\_L\_HH} &
$M_{rb}\geq51.76$ kmol & $M_{rb}<49.2$ kmol &
Stop feed, reactor heating, and reactor outlet & Yes\\

I7 & \id{RB\_L\_LL} &
$M_{rb}\leq12.08$ kmol & $M_{rb}>12.7$ kmol &
Stop bottoms withdrawal and reboiler steam & No\\

I8 & \id{DR\_L\_HH} &
$M_d\geq45.37$ kmol & $M_d<43.1$ kmol &
Stop feed, reactor heating, reactor outlet, and steam & Yes\\

I9 & \id{DR\_L\_LL} &
$M_d\leq10.59$ kmol & $M_d>11.1$ kmol &
Stop feed, reactor heating, reactor outlet, reflux, distillate, and steam & Yes\\

I10 & \id{COL\_P\_HH} &
$P_c\geq1.35$ bar & $P_c<1.28$ bar &
Stop feed, reactor heating, reactor outlet, and steam; fully open cooling water & Yes\\

I11 & \id{COL\_L\_LL} &
$M_{10}\leq0.30$ kmol & $M_{10}>0.315$ kmol &
Stop feed, reactor heating, reactor outlet, and steam & Yes\\
\bottomrule
\end{tabularx}
\end{table*}

The protective responses are defined for the benchmark rather than obtained from a formal hazard and operability study or safety integrity level determination. In particular, the reactor-pressure protection uses two stages. At the first high-high pressure threshold, heating is removed and gas discharge capacity is maximised while fresh feed is maintained. This configuration preserves the feed-associated thermal load of the endothermic reactor. At the escalated high-high-high threshold, feed is additionally restricted and a plant-stop event is latched. The thresholds and override actions are therefore properties of the simulated benchmark, not recommended settings for an industrial installation.

The protective logic is implemented independently of the LLM decision workflow. The agents cannot modify trip thresholds, suppress the protective calculations, or bypass their commanded overrides. Nevertheless, an imposed protective command does not guarantee the corresponding physical response. For example, an interlock may demand that PCV-101 open fully, while a vent-capacity fault prevents the actual discharge capacity from increasing to the requested level. The benchmark thus distinguishes correct protective decision logic from successful execution by the affected equipment.

\subsection{Supervisory authority and command arbitration}
\label{sec:arbitration}

The interface between supervisory decisions and the physical process is explicitly defined. For each manipulated input, the simulator distinguishes four successive command stages:
\begin{equation}
\begin{aligned}
&\underbrace{u^{\mathrm{BPCS}}}_{\text{regulatory request}} \longrightarrow \underbrace{u^{\mathrm{sup}}}_{\text{supervisory arbitration}} \\
&\quad \longrightarrow \underbrace{u^{\mathrm{SIS}}}_{\text{protective arbitration}} \longrightarrow \underbrace{u^{\mathrm{act}}}_{\text{plant input}}.
\end{aligned}
\label{eq:command_chain}
\end{equation}
The BPCS determines regulatory commands from process measurements and controller setpoints. Supervisory requests can change permitted setpoints, production throughput, standby-equipment states, and auxiliary equipment operation. The protective layer then applies any active interlock overrides before the commanded value reaches the plant. Finally, actuator or equipment faults determine whether the final element can deliver the requested action.

Conceptually, this ordering can be represented as
\begin{align}
u^{\mathrm{sup}} &=
\mathcal A_{\mathrm{sup}}
\left(u^{\mathrm{BPCS}},a\right),\\
u^{\mathrm{SIS}} &=
\mathcal A_{\mathrm{SIS}}
\left(u^{\mathrm{sup}},q\right),\\
u^{\mathrm{act}} &=
\mathcal F
\left(u^{\mathrm{SIS}},f\right),
\label{eq:command_arbitration}
\end{align}
where $a$ denotes the permitted supervisory request, $q$ the active interlock states, and $f$ the equipment and actuator fault conditions. The maps $\mathcal A_{\mathrm{sup}}$, $\mathcal A_{\mathrm{SIS}}$, and $\mathcal F$ represent supervisory command processing, protective arbitration, and physical fault effects, respectively. Equation~\eqref{eq:command_arbitration} describes the order of authority and is not an additional dynamic model.

This structure imposes two important constraints on autonomous operation. First, supervisory interventions are restricted to the operator action interface; an agent cannot directly assign arbitrary values to physical plant inputs or alter fault variables. Second, even an accepted and executable action remains subordinate to active protective commands and physical equipment limitations.

An illustrative example is a reduction in reactor outlet capacity. The reactor-level controller may increase its requested outlet opening, while the capacity fault in the valve restricts the actual flow. A supervisory setpoint adjustment may then reduce the rate of inventory accumulation, but it cannot make the physical fault in the plant go away. Similarly, starting a healthy standby pump can recover flow after a duty-pump trip, whereas commanding a failed valve to open further may have no effect. These distinctions allow the benchmark to evaluate the logical validity of proposed operator actions and their consequences in the simulated plant.

The detailed supervisory action grammar, including permitted command types and action validation, is introduced in Section~\ref{sec:framework}.

\subsection{Piping and instrumentation representation}

\begin{figure*}[pos=ht]
\centering
\includegraphics[width=\textwidth]{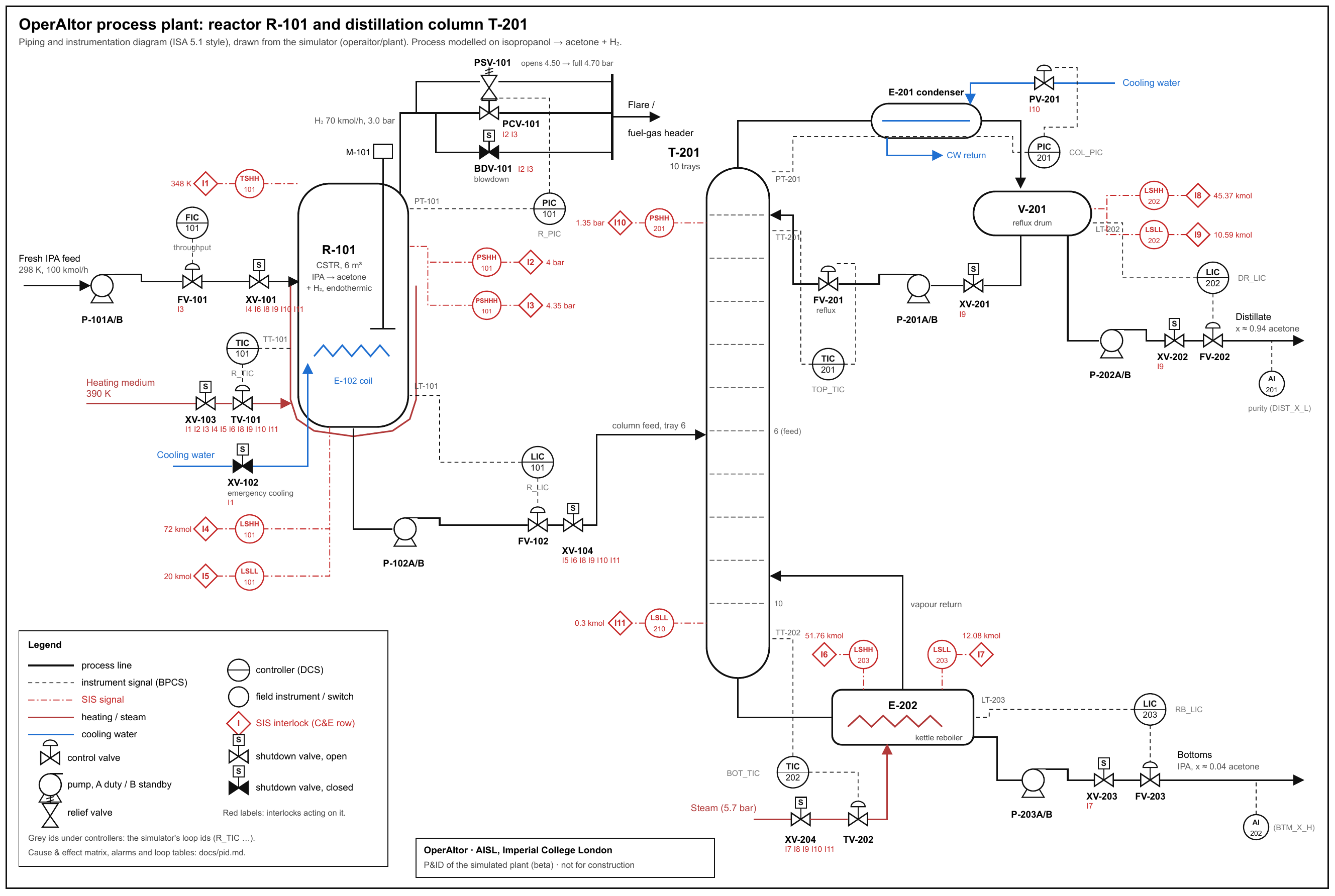}
\caption{Piping and instrumentation diagram of the integrated reactor--distillation benchmark. Dashed grey lines represent BPCS signals, dash-dotted red lines represent simulated protective signals, and numbered diamonds identify interlocks I1--I11.}
\label{fig:pid}
\end{figure*}

Figure~\ref{fig:pid} presents the benchmark piping and instrumentation diagram (P\&ID), using instrumentation conventions based on ISA-5.1 \citep{isa51}. The diagram identifies the process equipment, measurement points, eight regulatory control loops, final elements, protective switches, and eleven interlocks. Regulatory control signals and protective command paths are represented separately to distinguish the BPCS and SIS responsibilities.

The P\&ID and cause-and-effect documentation are generated from the simulator's equipment and protective-logic definitions. Automated consistency checks compare the documented interlock behaviour with the implemented logic, reducing the possibility of discrepancies between the benchmark description and the executable simulator. The software also provides an editable diagram and an interactive P\&ID view through the operator console.

Together, the regulatory, alarm, and protective layers establish the operational environment in which the agentic framework is evaluated. The following section describes how OperAItor uses process observations and plant knowledge to generate supervisory actions, and how alternative validation mechanisms determine which proposals are dispatched to this interface.
\section{OperAItor: an agentic framework for supervisory fault recovery}
\label{sec:framework}

\subsection{Architecture and execution workflow}
\label{sec:agent_architecture}

OperAItor is an agentic supervisory framework that converts process observations into operator-level interventions during abnormal operation. It operates above the BPCS and simulated SIS introduced in Section~\ref{sec:control}: regulatory controllers continue to regulate the process, protective interlocks retain priority over commanded actions, and the agent is restricted to the supervisory interface. The objective is not to replace these layers, but to connect fault evidence with an executable response when regulatory control alone cannot restore acceptable operation.

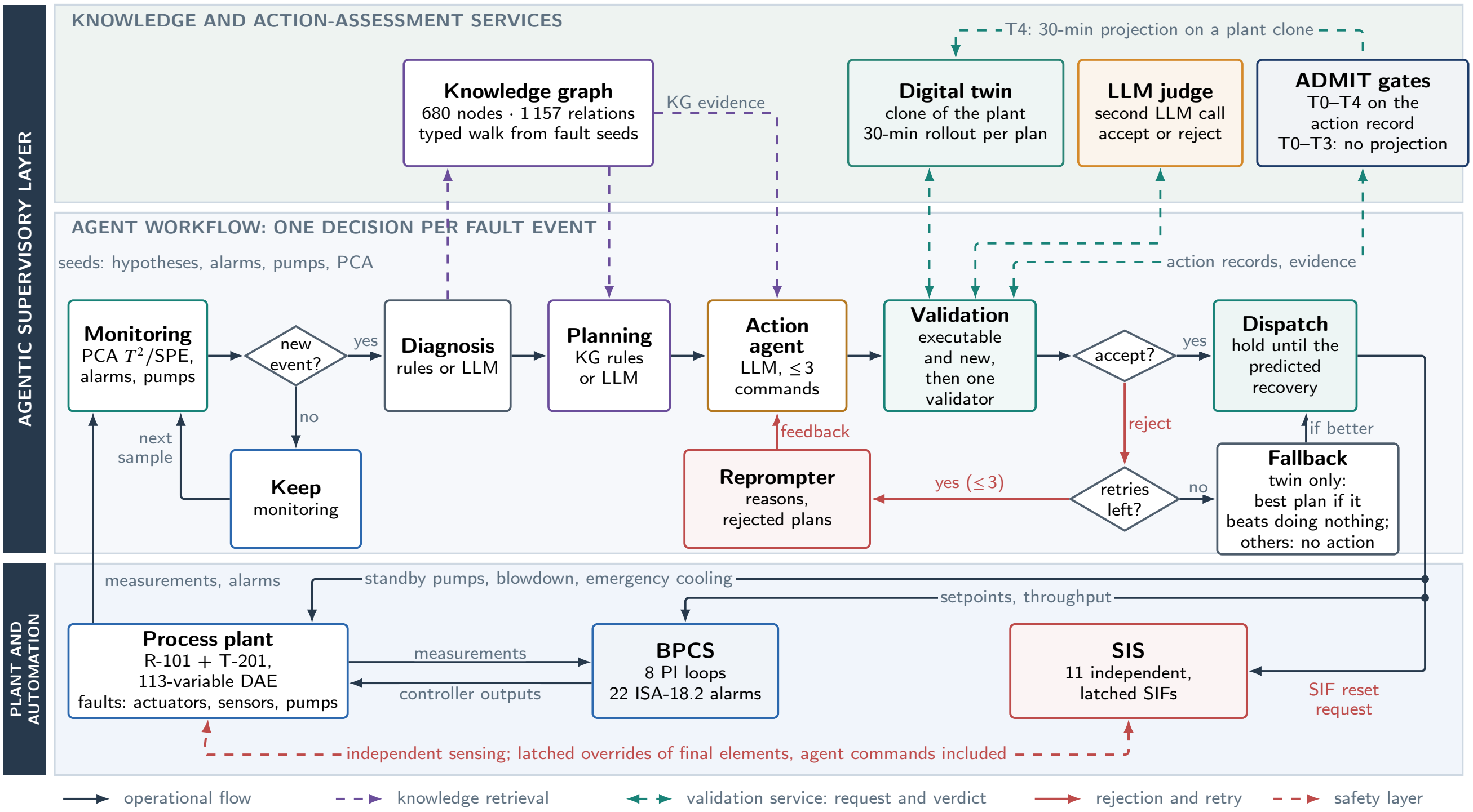
\begin{figure*}[tbp]
\centering
\definecolor{opInk}{HTML}{27394C}%
\definecolor{opMuted}{HTML}{5E7080}%
\definecolor{opBorder}{HTML}{CED8E2}%
\definecolor{opServiceBg}{HTML}{EEF3F1}%
\definecolor{opAgentBg}{HTML}{F6F8FB}%
\definecolor{opPhysicalBg}{HTML}{F1F5FB}%
\definecolor{opBlue}{HTML}{2C68AE}%
\definecolor{opGreen}{HTML}{198378}%
\definecolor{opViolet}{HTML}{6B4E9B}%
\definecolor{opOrange}{HTML}{B47820}%
\definecolor{opRed}{HTML}{BE4B4A}%
\definecolor{cTwin}{HTML}{2A8576}%
\definecolor{cAdmit}{HTML}{1D3B69}%
\definecolor{cJudge}{HTML}{D4892A}%
\tikzset{%
  op/.style={font=\sffamily},%
  opBox/.style={draw=opInk!82,line width=0.7pt,fill=white,rounded corners=1.6pt,align=center,%
       inner xsep=2.5pt,inner ysep=2.2pt,minimum height=1.3cm,font=\sffamily\fontsize{5.9}{6.9}\selectfont,%
       execute at begin node={\hyphenpenalty=10000}},%
  opServiceBox/.style={opBox,minimum height=1.25cm},%
  opPhysBox/.style={opBox,minimum height=1.1cm},%
  opDecision/.style={diamond,aspect=1.7,draw=opInk!80,line width=.7pt,fill=white,align=center,%
       inner sep=0pt,font=\sffamily\fontsize{6}{7}\selectfont},%
  opFlow/.style={draw=opInk,line width=.7pt,-{Latex[length=1.6mm,width=1.2mm]},rounded corners=1.2pt},%
  opRetrieval/.style={draw=opViolet,line width=.7pt,dashed,-{Latex[length=1.6mm,width=1.2mm]},rounded corners=1.2pt},%
  opValidation/.style={draw=opGreen,line width=.7pt,dashed,{Latex[length=1.6mm,width=1.2mm]}-{Latex[length=1.6mm,width=1.2mm]},%
       rounded corners=1.2pt},%
  opRetry/.style={draw=opRed,line width=.75pt,-{Latex[length=1.6mm,width=1.2mm]},rounded corners=1.2pt},%
  opSafety/.style={draw=opRed,line width=.75pt,dashed,-{Latex[length=1.6mm,width=1.2mm]},rounded corners=1.2pt},%
  opLab/.style={font=\sffamily\fontsize{5.6}{6.4}\selectfont,text=opMuted,fill=#1,inner sep=0.8pt},%
  opLab/.default=opAgentBg,%
  opRedLab/.style={font=\sffamily\fontsize{5.6}{6.4}\selectfont,text=opRed,fill=#1,inner sep=0.8pt},%
  opRedLab/.default=opAgentBg,%
  opBand/.style={font=\sffamily\bfseries\fontsize{5.6}{6.4}\selectfont,text=opMuted,anchor=west},%
}%
\providecommand{\opT}[1]{{\fontsize{7}{8.2}\selectfont\bfseries #1}}%
\providecommand{\opD}[1]{\\#1}%
\begin{tikzpicture}[x=1cm,y=1cm,op]

\def\R{17.12}
\fill[opServiceBg]  (0.6,7.1) rectangle (\R,9.42);
\fill[opAgentBg]    (0.6,2.98) rectangle (\R,6.98);
\fill[opPhysicalBg] (0.6,0.38) rectangle (\R,2.86);
\foreach \a/\b in {7.1/9.42,2.98/6.98,0.38/2.86} \draw[opBorder,line width=.6pt] (0.6,\a) rectangle (\R,\b);
\fill[opInk] (0.0,2.98) rectangle (0.5,9.42);
\fill[opInk] (0.0,0.38) rectangle (0.5,2.86);
\node[rotate=90,text=white,font=\sffamily\bfseries\fontsize{6.4}{7}\selectfont] at (0.25,6.2) {AGENTIC SUPERVISORY LAYER};
\node[rotate=90,text=white,font=\sffamily\bfseries\fontsize{5.4}{6}\selectfont,align=center] at (0.25,1.62) {PLANT AND\\AUTOMATION};
\node[opBand] at (0.68,9.24) {KNOWLEDGE AND ACTION-ASSESSMENT SERVICES};
\node[opBand] at (0.68,6.81) {AGENT WORKFLOW: ONE DECISION PER FAULT EVENT};

\def\gap{0.42}
\def\yMain{5.3}
\def\yLow{3.62}
\node[opBox,draw=opGreen,text width=1.45cm,anchor=west] (mon) at (0.75,\yMain) {\opT{Monitoring}\opD{PCA $T^2$/SPE,}\opD{alarms, pumps}};
\node[opDecision,text width=0.62cm,minimum width=1.1cm,anchor=west] (event) at ($(mon.east)+(\gap,0)$) {new\\event?};
\node[opBox,text width=1.3cm,anchor=west] (diag) at ($(event.east)+(\gap,0)$) {\opT{Diagnosis}\opD{rules or LLM}};
\node[opBox,draw=opViolet,text width=1.25cm,anchor=west] (plan) at ($(diag.east)+(\gap,0)$) {\opT{Planning}\opD{KG rules or LLM}};
\node[opBox,draw=opOrange,text width=1.45cm,anchor=west] (action) at ($(plan.east)+(\gap,0)$) {\opT{Action agent}\opD{LLM, $\leq$\,3 commands}};
\node[opBox,draw=opGreen,fill=opGreen!6,text width=1.6cm,anchor=west] (valid) at ($(action.east)+(\gap,0)$)
  {\opT{Validation}\opD{executable and new,}\opD{then one validator}};
\node[opDecision,text width=0.7cm,minimum width=1.2cm,anchor=west] (accept) at ($(valid.east)+(\gap,0)$) {accept?};
\node[opBox,draw=opGreen,fill=opGreen!8,text width=1.5cm,anchor=west] (dispatch) at ($(accept.east)+(\gap,0)$)
  {\opT{Dispatch}\opD{hold until the}\opD{predicted recovery}};

\node[opBox,draw=opBlue,text width=1.35cm,minimum height=1.15cm] (continue) at (event |- 0,\yLow) {\opT{Keep}\opD{monitoring}};
\node[opBox,draw=opRed,fill=opRed!6,text width=2.0cm,minimum height=1.15cm] (repro) at (action |- 0,\yLow)
  {\opT{Reprompter}\opD{reasons, rejected plans}};
\node[opDecision,text width=0.62cm,minimum width=1.1cm] (retryq) at (accept |- 0,\yLow) {retries\\left?};
\node[opBox,text width=1.95cm,minimum height=0.8cm,anchor=west] (fallback) at ($(retryq.east)+(\gap,0)$)
  {\opT{Fallback}\opD{twin only: best plan if it}\opD{beats doing nothing;}\opD{others: no action}};
\coordinate (trunk) at ($(fallback.east)+(0.3,0)$);

\node[opServiceBox,draw=opViolet,text width=2.75cm] (kg) at ($($(diag)!0.5!(plan)$)+(0,2.85)$)
  {\opT{Knowledge graph}\opD{680 nodes $\cdot$ 1\,157 relations}\opD{typed walk from fault seeds}};
\node[opServiceBox,draw=cTwin,fill=cTwin!8,text width=2.35cm] (twin) at ($(valid)+(-0.05,2.85)$)
  {\opT{Digital twin}\opD{clone of the plant}\opD{30-min rollout per plan}};
\node[opServiceBox,draw=cJudge,fill=cJudge!9,text width=1.75cm,anchor=west] (judge) at ($(twin.east)+(0.15,0)$)
  {\opT{LLM judge}\opD{second LLM call}\opD{accept or reject}};
\node[opServiceBox,draw=cAdmit,fill=cAdmit!6,text width=2.3cm,anchor=west] (admit) at ($(judge.east)+(0.15,0)$)
  {\opT{ADMIT gates}\opD{T0--T4 on the action record}\opD{T0--T3: no projection}};
\draw[opValidation,{Latex[length=1.6mm,width=1.2mm]}-] (twin.north) -- (twin.north |- 0,9.12) -- (admit.north |- 0,9.12) -- (admit.north);
\node[opLab=opServiceBg] at ($(twin.north |- 0,9.12)!0.5!(admit.north |- 0,9.12)$) {T4: 30-min projection on a plant clone};

\draw[opRetrieval] (diag.north) -- (diag.north |- kg.south);
\node[opLab,anchor=east] at ($(diag.west |- 0,6.38)+(-0.08,0)$) {seeds: hypotheses, alarms, pumps, PCA};
\draw[opRetrieval] (plan.north |- kg.south) -- (plan.north);
\draw[opRetrieval] (kg.east) -- (action.north |- kg.east) -- (action.north);
\node[opLab=opServiceBg,anchor=south] at ($(kg.east)!0.5!(action.north |- kg.east)+(0,0.02)$) {KG evidence};

\coordinate (vT) at ($(valid.north west)!0.3!(valid.north east)$);
\coordinate (vJ) at ($(valid.north west)!0.6!(valid.north east)$);
\coordinate (vA) at ($(valid.north west)!0.85!(valid.north east)$);
\draw[opValidation] (vT) -- (vT |- twin.south);
\draw[opValidation] (vJ) -- (vJ |- 0,6.62) -- (judge.south |- 0,6.62) -- (judge.south);
\draw[opValidation] (vA) -- (vA |- 0,6.4) -- (admit.south |- 0,6.4) -- (admit.south);
\node[opLab] at ($(judge.south |- 0,6.4)!0.5!(admit.south |- 0,6.4)$) {action records, evidence};

\draw[opFlow] (mon.east) -- (event.west);
\draw[opFlow] (event.east) -- node[opLab,above=1pt] {yes} (diag.west);
\draw[opFlow] (diag.east) -- (plan.west);
\draw[opFlow] (plan.east) -- (action.west);
\draw[opFlow] (action.east) -- (valid.west);
\draw[opFlow] (valid.east) -- (accept.west);
\draw[opFlow] (accept.east) -- node[opLab,above=1pt] {yes} (dispatch.west);
\draw[opFlow] (event.south) -- node[opLab,right] {no} (continue.north);
\coordinate (loopx) at ($(mon.south east)+(-0.32,0)$);
\draw[opFlow] (continue.west) -- (loopx |- continue.west) -- (loopx);
\node[opLab,anchor=east,align=right] at ($(loopx |- 0,4.2)+(-0.06,0)$) {next\\sample};

\draw[opRetry] (accept.south) -- node[opRedLab,right] {reject} (retryq.north);
\draw[opRetry] (retryq.west) -- node[opRedLab,above=1pt] {yes ($\leq$\,3)} (repro.east);
\draw[opRetry] (repro.north) -- node[opRedLab,right] {feedback} (action.south);
\draw[opFlow] (retryq.east) -- node[opLab,above=1pt] {no} (fallback.west);
\coordinate (fbx) at ($(dispatch.south west)!0.5!(fallback.north east |- dispatch.south)$);
\draw[opFlow] (fbx |- fallback.north) -- node[opLab,right] {if better} (fbx);

\node[opPhysBox,draw=opBlue,text width=3.1cm] (plant) at (2.4,1.6)
  {\opT{Process plant}\opD{R-101 + T-201, 113-variable DAE}\opD{faults: actuators, sensors, pumps}};
\node[opPhysBox,draw=opBlue,fill=opBlue!7,text width=2.0cm] (bpcs) at (8.0,1.6) {\opT{BPCS}\opD{8 PI loops}\opD{22 ISA-18.2 alarms}};
\node[opPhysBox,draw=opRed,fill=opRed!6,text width=2.6cm] (sis) at (13.2,1.6)
  {\opT{SIS}\opD{11 independent,}\opD{latched SIFs}};

\draw[opFlow] (plant.north -| 1.05,0) -- (mon.south -| 1.05,0);
\node[opLab=opPhysicalBg,anchor=west] at (1.15,2.66) {measurements, alarms};
\draw[opFlow] ([yshift=3pt]plant.east) -- node[opLab=opPhysicalBg,above] {measurements} ([yshift=3pt]bpcs.west);
\draw[opFlow] ([yshift=-4pt]bpcs.west) -- node[opLab=opPhysicalBg,below] {controller outputs} ([yshift=-4pt]plant.east);
\draw[opFlow] (dispatch.east) -- (trunk |- dispatch.east) -- (trunk |- sis.east) -- (sis.east);
\fill[opInk] (trunk |- 0,2.46) circle (1.3pt);
\fill[opInk] (trunk |- 0,2.68) circle (1.3pt);
\draw[opFlow] (trunk |- 0,2.46) -- (bpcs.north |- 0,2.46) -- (bpcs.north);
\node[opLab=opPhysicalBg] at (12.0,2.46) {setpoints, throughput};
\draw[opFlow] (trunk |- 0,2.68) -- (3.62,2.68) -- (plant.north -| 3.62,0);
\node[opLab=opPhysicalBg] at (6.4,2.68) {standby pumps, blowdown, emergency cooling};
\node[opRedLab=opPhysicalBg,align=center,anchor=north] at ($(trunk |- 0,1.5)+(-0.95,0)$) {SIF reset\\request};
\draw[opSafety,{Latex[length=1.6mm,width=1.2mm]}-{Latex[length=1.6mm,width=1.2mm]}] (sis.south) -- (13.2,0.62) -- (2.35,0.62) -- (plant.south);
\node[opRedLab=opPhysicalBg] at (7.9,0.62) {independent sensing; latched overrides of final elements, agent commands included};

\begin{scope}[shift={(0.7,0.1)},every node/.style={anchor=west,font=\sffamily\fontsize{5.6}{6.4}\selectfont,text=opMuted}]
  \draw[opFlow] (0,0) -- (0.55,0);         \node at (0.6,0) {operational flow};
  \draw[opRetrieval] (3.2,0) -- (3.75,0);  \node at (3.8,0) {knowledge retrieval};
  \draw[opValidation] (6.6,0) -- (7.15,0); \node at (7.2,0) {validation service: request and verdict};
  \draw[opRetry] (11.4,0) -- (11.95,0);    \node at (12.0,0) {rejection and retry};
  \draw[opSafety] (14.2,0) -- (14.75,0);   \node at (14.8,0) {safety layer};
\end{scope}
\end{tikzpicture}
\caption{OperAItor supervisory workflow. Plant observations support monitoring, diagnosis, knowledge-graph retrieval, planning and LLM action generation. Proposed actions are screened for executability and passed to the selected validation layer before dispatch through the command-arbitration hierarchy of Section~\ref{sec:arbitration}. Rejected proposals can be returned for revision.}
\label{fig:framework}
\end{figure*}

Figure~\ref{fig:framework} illustrates the workflow, implemented as a LangGraph state machine \citep{langgraph}. At a decision point, the monitoring module supplies abnormality statistics and process trends; the diagnosis module forms fault hypotheses; knowledge-graph retrieval identifies associated equipment, constraints and candidate interventions; the planning module constructs an operational brief; and an LLM action agent proposes a structured plan. The plan then passes through a common executability check and a selected action-validation mechanism before dispatch. When a proposal is rejected, the validation outcome is returned to the action agent through a reprompting edge. Section~\ref{sec:validation} describes the validators and their retry and fallback rules; the present section describes the common proposal-generation architecture. Configuration-specific choices of deterministic or LLM planning are reported with the evaluation protocol.


The plant simulator and the agent workflow execute as separate asynchronous tasks. For the comparative benchmark, simulated process time is held fixed during each agent decision. Consequently, LLM inference and validation latency contribute to measured wall-clock decision cost but do not allow the simulated plant to evolve before an action is applied. The software also permits asynchronous execution with a continuously advancing plant clock, but that operating mode is not the basis of the results reported here. This distinction avoids interpreting benchmark outcomes as demonstrations of real-time decision-making under communication and inference delays.

\subsection{Process monitoring and decision triggers}
\label{sec:pca}

The monitoring module uses principal component analysis (PCA) to identify departures from normal operating behaviour \citep{macgregor1995,qin2003}. Its input vector contains 42 features: 34 process measurements covering reactor and column temperatures, pressures, inventories and compositions, including the ten tray temperatures and pressures, together with the outputs of the eight regulatory controllers. Including controller outputs allows the monitoring system to observe compensatory control effort as well as deviations in the measured process variables.

The PCA model is fitted to 3\,040 samples collected at 0.5-min intervals during normal-operation simulations. These include nominal operation, three runs with randomised disturbance steps, steady operation at 70\% and 85\% throughput, and positive and negative variations of the configured disturbances. For a standardised observation $\mathbf z=(\mathbf x-\bar{\mathbf x})\oslash\mathbf s$, retained loading matrix $\mathbf P$ and diagonal eigenvalue matrix $\boldsymbol\Lambda$, the monitoring statistics are
\begin{align}
\mathbf t &= \mathbf z\mathbf P,\\
T^2 &= \mathbf t\boldsymbol\Lambda^{-1}\mathbf t^\top,\\
Q &= \big\|\mathbf z-\mathbf t\mathbf P^\top\big\|_2^2,
\label{eq:monitor_statistics}
\end{align}
where $T^2$ measures variation within the retained PCA subspace and $Q$, also termed the squared prediction error (SPE), measures the residual outside that subspace. The model retains 23 components, explaining 95.6\% of the training-set variance. For each statistic, the detection threshold is selected as the larger of its theoretical 99th-percentile control limit and the empirical 99th percentile of the training statistic. The resulting thresholds are $T^2_{\mathrm{lim}}=67.9$ and $Q_{\mathrm{lim}}=5.07$; the corresponding theoretical limits are 42.1 and 4.78, respectively \citep{hotelling1947,jackson1979,nomikos1995}. The empirical comparison accommodates departures from Gaussian assumptions in the disturbance-containing training data.

To support fault attribution, the monitor ranks variables using their individual contributions to the two statistics:
\begin{align}
c_i^{T^2} &= z_i\big(\mathbf P\boldsymbol\Lambda^{-1}
\mathbf P^\top\mathbf z^\top\big)_i,\\
c_i^Q &= \big(z_i-(\mathbf t\mathbf P^\top)_i\big)^2,\\
s_i &= \frac{\max(c_i^{T^2},0)}{T^2_{\mathrm{lim}}}
     +\frac{c_i^Q}{Q_{\mathrm{lim}}}.
\label{eq:pca_contributions}
\end{align}
A fault is confirmed when at least one monitoring statistic exceeds its threshold for three consecutive samples. In the benchmark's initial decision protocol, PCA confirmation is the decision trigger; an alarm, pump-trip indication or saturated controller does not independently trigger an agent decision. These signals nevertheless enter the subsequent diagnosis as evidence. The monitor also supplies local trends and a linear extrapolation of time to an alarm or protective limit, $\tau=|L-x|/|\dot x|$ for a non-zero estimated rate directed towards the limit. This quantity is a short-horizon diagnostic indicator, not a validated prediction of trip time. Decision triggering and the episode timing protocol are specified in Section~\ref{sec:benchmark}.

\subsection{Fault diagnosis and evidence assembly}
\label{sec:diagnosis}

The diagnosis stage translates statistical deviations and operational signals into ranked fault hypotheses. OperAItor provides two alternatives. A deterministic heuristic compares the highest PCA contributions, active alarms, failed pump indications and saturated loops against the symptom signatures in the fault catalogue. Alternatively, an LLM diagnosis agent receives the same operating evidence together with the fault catalogue and returns a structured ranking of candidate faults.

The LLM diagnoser is given the current $T^2$ and $Q$ values relative to their limits, the variables contributing most strongly to each statistic, active alarms, controller saturation, pump states, measured trends and estimated times to relevant limits. It is constrained to hypotheses named in the catalogue and is asked to identify the fault class (sensor, actuator, equipment or process), evidence supporting and contradicting each hypothesis, and checks an operator could perform to discriminate among them. Hypotheses are therefore accompanied by an explicit evidence trail rather than being passed downstream as unsupported fault labels.

A diagnosis remains an inference, not a verified statement about the injected fault. In particular, an incorrect or incomplete diagnosis can affect which interventions are considered by the planner. The action-validation mechanisms are evaluated downstream of this imperfect information pipeline. In the matched-model digital-twin configuration, however, the rollout uses a clone of the true simulated plant state; this oracle information is a specific benchmark assumption discussed in Section~\ref{sec:validation}.

\subsection{Process knowledge graph and recovery planning}
\label{sec:kg_planning}

OperAItor represents plant-specific knowledge in a process knowledge graph (KG) \citep{hogan2021kg}. The graph comprises 680 nodes and 1\,157 relations, serialised as JSON-LD/TriG and organised using concepts from the VDI~3682 formalised process description and the SOSA sensor/actuator ontology \citep{vdi3682,janowicz2019sosa}. It encodes equipment connectivity, 40 observable and 11 actuatable process properties, eight controllers, five pump trains, 22 alarms and eleven protective interlocks with 37 override rules. The graph additionally includes 16 represented fault modes linked to 20 symptoms and operator actions linked to the equipment or variables they affect.

The KG preserves the distinction between requested, supervisory, safety-arbitrated and actual actuator signals introduced in Section~\ref{sec:arbitration}. This is important when a fault changes the physical response to an otherwise valid command. For example, a controller and a protective interlock may both demand additional hydrogen venting without being able to overcome a vent-capacity restriction. A translation layer links graph identifiers to the corresponding simulator entities and legal supervisory commands, making retrieved relationships actionable rather than merely descriptive.

At each decision, retrieval begins with the three highest-ranked diagnostic hypotheses, active alarms, saturated loops, failed pump trains and the most influential PCA variables. A typed graph traversal expands one hop in both directions over relevant equipment, property, fault and protection relationships. Actions connected to retrieved entities through an \emph{actsOn} relation are then added to the candidate context. This provides a bounded, plant-specific form of knowledge-grounded retrieval rather than passing the entire graph to the LLM \citep{lewis2020rag,pan2024kgllm}.

A planning module converts the diagnosis and retrieved context into an operational brief comprising the suspected fault, supporting symptoms, recovery objective, feasible supervisory levers and constraints on their use. Either a deterministic KG-based planner or an LLM planner may generate this brief from the same evidence. For example, a diagnosed reflux duty-pump trip can retrieve the associated standby train and the \id{START\_SPARE} command; a saturated column-pressure controller instead suggests that further setpoint or valve demands may not deliver the intended process response. The planner supplies candidate actions and their rationale, but does not issue commands to the plant. The benchmark configuration of planner and diagnoser is documented with the experimental protocol in Section~\ref{sec:benchmark}.

\subsection{Action generation and supervisory interface}
\label{sec:action_agent}

The action agent receives the current plant snapshot, the ranked diagnostic hypotheses, the planning brief, the retrieved KG evidence, the controller directions and operating limits, and a machine-readable action grammar. It returns a JSON plan with at most three commands and a justification. A plan can combine interventions, for example starting a standby pump while adjusting a supervisory target. Every command belongs to the permitted interface in Table~\ref{tab:action_interface}. All commands in the workflow come from the LLM while the validation and dispatch stages decide whether they are accepted and executed.

\begin{table}[t]
\centering
\caption{Supervisory action interface exposed to the OperAItor action agent. Reset requests and alarm acknowledgements are additional interface operations, rather than direct plant-input assignments.}
\label{tab:action_interface}
\small
\begin{tabularx}{\linewidth}{@{}p{4.0cm}X@{}}
\toprule
Command & Permitted effect\\
\midrule
\id{SET\_SETPOINT} & Change the target of one of the eight PI loops within the range in Table~\ref{tab:loops}.\\
\id{SET\_THROUGHPUT} & Change the fresh-feed production target through FIC-101.\\
\id{START\_SPARE} & Start the available standby pump of a specified duty/standby train.\\
\id{OPERATE\_AUXILIARY} & Request a permitted auxiliary operation, such as hydrogen blowdown or emergency cooling.\\
\id{NO\_ACTION} & Make no supervisory change at the current decision point.\\
\bottomrule
\end{tabularx}
\end{table}

The interface also supports alarm acknowledgement and requests to reset eligible latched non-trip interlocks once their reset conditions are satisfied. It does not permit the agent to change PI tuning parameters, write physical plant inputs, manipulate fault or disturbance variables, change protection thresholds or bypass the SIS. Thus, a proposed intervention may be operationally meaningful without granting the agent unrestricted access to the simulation state.

Before reaching the selected validator, a plan undergoes an executability check against the action schema and the current interface constraints. Unknown action types or targets, out-of-range setpoints and malformed requests are rejected and returned for correction. Passing this check means only that the request can be represented and dispatched; it does not establish that the action will recover the process or avoid a protective trip. That distinction motivates the validation methods examined in Section~\ref{sec:validation}.

For the benchmark, inference settings and model choices are held fixed within each experimental configuration. GPT-5-family reasoning models are called with low reasoning effort, while locally served open models use Ollama at temperature zero \citep{ollama}. These implementation settings affect reproducibility and decision cost but do not alter the supervisory authority described above.

In summary, OperAItor establishes a common path from evidence of abnormal operation to candidate operator actions. Monitoring and diagnosis determine what the agent believes has happened; KG retrieval and planning determine which interventions appear relevant; and the action agent converts those interventions into executable requests. The following section compares mechanisms for deciding whether these requests should be applied at the same supervisory interface; configuration-specific choices of planner are made explicit in the benchmark protocol.

\section{Action validation methods}
\label{sec:validation}

The principal experimental variable in this study is the mechanism used to evaluate an LLM-generated operator action before it is dispatched to the simulated process. OperAItor separates action generation from validation: the common agent workflow described in Section~\ref{sec:framework} produces an operator-level proposal, and a selected validation configuration determines whether the proposal may proceed through the supervisory command interface. The regulatory controllers and protective interlocks remain active irrespective of the selected validator. Thus, admission by a validator does not bypass the final command arbitration described in Section~\ref{sec:arbitration}.

The validation problem is distinct from syntactic executability. A command may refer to a known controller, respect its setpoint range, and be accepted by the interface while still worsening a process inventory or accelerating a protective trip. Conversely, rejecting every executable command would prevent agent-induced changes but also eliminate opportunities for recovery.

We therefore examine five configurations that impose different requirements on a proposed intervention: executability only, review by a second LLM, prediction with a digital twin, deterministic ADMIT admissibility checks, and ADMIT checks supplemented by consequence projection. Table~\ref{tab:validators} summarises their information requirements, acceptance criteria, and fallback behaviour.

\subsection{Common proposal and validation interface}
\label{sec:validator_interface}

At decision point $k$, the action agent receives the operating evidence assembled by the preceding modules and submits a proposal $a_k$ containing at most three commands from Table~\ref{tab:action_interface}. Each proposal includes a justification; ADMIT configurations additionally require structured action records as described below. Before reaching any of the five validators, a common executability check verifies the command types and targets, parameter formats, permitted setpoint ranges, and current interface constraints. Proposals that fail this check are rejected without being dispatched. Passing executability establishes only that the request can be represented at the operator interface, not that it is appropriate for the current fault.

The selected validator receives the proposal and the information available to that configuration. This information may include the current plant snapshot, alarm and interlock states, process trends, diagnostic hypotheses, the planning brief, and retrieved knowledge-graph (KG) context. Importantly, the configurations do not have identical information requirements: the LLM judge reviews the supplied evidence, ADMIT evaluates an action record against a trusted ledger and safety case, and the digital twin evaluates predicted process trajectories. A rejection is returned to the action agent with feedback that may support a revised proposal. At most four proposals are permitted within one decision, comprising the initial proposal and up to three subsequent attempts. The configuration-specific behaviour when these attempts are exhausted is described below.

\begin{table*}[t]
\centering
\caption{Action-validation configurations. All proposals pass the common executability check before the listed validation mechanism. A ``no action'' fallback means that no new supervisory command is dispatched at that decision point; existing BPCS and protective functions continue to operate.}
\label{tab:validators}
\small
\renewcommand{\arraystretch}{1.12}
\begin{tabularx}{\textwidth}{@{}
>{\raggedright\arraybackslash}p{2.6cm}
>{\raggedright\arraybackslash}p{2.4cm}
>{\raggedright\arraybackslash}X
>{\raggedright\arraybackslash}p{3.0cm}@{}}
\toprule
Configuration & Additional requirement & Decision criterion & After unsuccessful attempts \\
\midrule
No validation & None &
Dispatch the first executable proposal without evaluating its operational consequences. &
No validation-specific fallback. \\

LLM judge & Second LLM call &
Accept when the judge approves the proposed intervention against written fault-relevance, direction, proportionality, and protective-limit criteria. &
No new action is dispatched. \\

Digital twin & 30-min process rollout &
Prefer a proposal that restores the active recovery targets without violating the protective envelope or creating new high-priority alarms; otherwise compare predicted mitigations against no action. &
Dispatch the best eligible mitigation only if it improves on the no-action projection. \\

ADMIT T0--T3 & Trusted evidence ledger and action safety case &
Check record integrity (T0), admissible evidence (T1), hazard representation (T2), and authority, operating mode, procedures, confidence, and reversibility (T3). &
No new action is dispatched. \\

ADMIT T0--T4 & ADMIT T0--T3 plus a 30-min rollout &
Apply T0--T3 and reject a projected protective-limit violation or a sufficient loss of safety margin relative to no action (T4). &
No new action is dispatched. \\
\bottomrule
\end{tabularx}
\end{table*}

\subsection{Executability-only and LLM-judge baselines}
\label{sec:baseline_validators}

The \emph{executability-only} configuration is the unvalidated baseline. It applies the first proposal that passes the common interface checks, without asking whether the plan addresses the diagnosed fault or whether its predicted effects are favourable. This establishes the outcome of the action-generation pipeline without an additional validation step. The absence of an explicit validator does not imply unrestricted control: the BPCS, supervisory action grammar, and protective arbitration remain in force.

In the \emph{LLM-judge} configuration, a second call to the same underlying model reviews the proposal before dispatch. The judge receives the candidate commands, current plant snapshot, diagnosis, planning brief, KG evidence, and command limits. Its written criteria ask whether the intervention addresses the suspected fault, respects control directions, avoids moving the process towards a protective limit, and is proportionate to the observed condition. The judge returns a structured decision containing \texttt{accept}, \texttt{confidence}, \texttt{reason}, \texttt{risks}, and \texttt{suggestion}. Only an accepted proposal is dispatched; a rejected proposal or failed judge call produces rejection feedback to the action agent. If no proposal is accepted within the permitted attempts, no new action is applied.

This baseline provides an example of model-based \emph{linguistic review}, rather than a physical consequence check: the judge reasons over the supplied description and observations but does not simulate the plant or independently verify an engineering safety case. Using the same model family for proposal and review may also leave the two calls vulnerable to correlated errors. Its measured cost includes the additional LLM inference and any proposals generated after rejection.

\subsection{Digital-twin consequence validation}
\label{sec:twin_validator}

The digital-twin validator tests candidate plans using a cloned state of the dynamic plant. Each candidate, including \id{NO\_ACTION}, is evaluated over a fixed horizon of $H=30$~min. The rollout begins from the simulator state at the decision point, preserves faults active at that instant, and excludes subsequently scheduled events, including future fault removal. The twin therefore estimates the consequences of continuing under the current fault conditions rather than relying on knowledge of when those faults will disappear. In the reported benchmark, the twin is an \emph{oracle matched-model clone}: it uses the same physical model and underlying state as the simulated plant, not a separately identified model with state-estimation or parameter uncertainty.

A candidate is classified as restoring operation when the recovery targets identified at the decision point return to their acceptable conditions and remain there for at least 2~min during the rollout, without violating the simulated protective envelope or producing a new priority-1 or priority-2 alarm. The targets include active process-alarm conditions, pump-related operating states, and, when PCA has confirmed the fault, the requirement that both $T^2$ and squared prediction error $Q$ return below their respective limits. The full-restoration criterion is stricter than avoiding an immediate trip: a proposal can keep the plant operating for 30~min without eliminating every fault signature.

When full restoration is not predicted, the validator compares candidate \emph{mitigations} with the \id{NO\_ACTION} rollout. Its lexicographic evaluation orders candidates using
\begin{align}
\boldsymbol\ell(a) =\big(&
\mathbb{1}[\text{trip}],\;-t_{\mathrm{trip}}(a),\;
N_{\mathrm{unmet}}(a),\notag\\
&f_{\mathrm{unsafe}}(a),\;J_{\mathrm{PCA}}(a)\big),
\label{eq:twin_rank}\\
J_{\mathrm{PCA}}(a) &=
\frac{T^2(a)}{T^2_{\mathrm{lim}}}
+\frac{Q(a)}{Q_{\mathrm{lim}}}.
\label{eq:twin_pca}
\end{align}
Here $t_{\mathrm{trip}}$ is the projected trip time when a trip occurs, $N_{\mathrm{unmet}}$ counts unresolved recovery targets, $f_{\mathrm{unsafe}}$ is the fraction of the projected trajectory outside the specified acceptable operating region, and the final term summarises the remaining normalised monitoring deviation. Non-tripping trajectories are preferred to tripping trajectories; among the latter, later trips are preferred. The tuple is compared lexicographically rather than combined into a weighted score. Terms involving trip time are relevant only when a trip is predicted.

After the available proposals have been considered, the best eligible mitigation may be dispatched if it improves on the no-action projection; otherwise no new action is applied. When a mitigation rather than full restoration is selected, the framework schedules a further decision after 10~min. This fallback is an important distinction between the digital twin and the other validators: a rejected full-recovery proposal does not necessarily imply that the plant receives no intervention. The ranking and mitigation rule are specific to the implemented benchmark and do not constitute a formal guarantee of constraint satisfaction beyond the modelled horizon.

\subsection{ADMIT admissibility and consequence gates}
\label{sec:admit_validator}

The two ADMIT configurations evaluate the \emph{action record} accompanying each proposed intervention rather than relying solely on the action-agent's explanation. They implement the Admissible Decisions for Machine Interventions (ADMIT) framework \citep{misra2026admitbench,admitbench_software}. The platform assembles an auditable, hash-chained ledger from the current measurements, PCA statistics, alarm and interlock states, pump conditions, and process trends. KG facts are also available as trusted context. LLM-generated diagnostic statements are treated as untrusted hypotheses and cannot, by themselves, satisfy a required evidence citation.

Each proposed command is associated with an action class, required authority, permitted operating modes, supporting evidence, applicable standard operating procedure (SOP) steps, confidence requirement, reversibility category, and any required recovery plan. The agent supplies a record referencing the evidence it relies on, the checks performed, the hazard addressed, and the rationale for the proposed intervention. For example, a supervisory setpoint change requires relevant process evidence and procedural checks that the loop is in automatic operation and its measurement is verified. Opening the reactor blowdown valve is treated as a higher-consequence intervention requiring pressure-related evidence and a recovery plan. Bypassing a protective interlock is not within the permissible operator action set.

The deterministic checks are organised as a sequence of gates:
\begin{description}
\item[T0: Record integrity.]
Verify the completeness and structural validity of the submitted action record and its references.

\item[T1: Evidence.]
Verify that cited evidence is present in the trusted ledger and satisfies the requirements associated with the proposed command.

\item[T2: Hazard.]
Verify that the record identifies the relevant hazard and provides the required hazard-related information.

\item[T3: Operational admissibility.]
Check command authority, current operating mode, required SOP steps, confidence and reversibility conditions, and recovery-plan requirements where applicable.

\item[T4: Consequence projection.]
When enabled, compare a 30-min rollout of the proposed action with the corresponding no-action projection and check projected protective-limit and safety-margin conditions.
\end{description}

The operating mode is classified as \emph{normal}, \emph{upset} (a PCA-confirmed fault or latched interlock), or \emph{emergency} (following a plant trip). Mode-specific permissions restrict which actions are admissible; for example, supervisory setpoint changes are not permitted in emergency mode. Gates T0--T3 evaluate the action's documented admissibility without predicting its physical effect. Consequently, an action can pass T0--T3 yet still have an undesirable process consequence. The T0--T3 configuration intentionally retains this limitation so that the contribution of consequence projection can be examined separately.

The T0--T4 configuration adds a 30-min model-based projection. Let $s_v$ be a monitored process variable with nominal value $s_v^{\mathrm{nom}}$ and relevant upper or lower protective limit $L_v$. A normalised margin is defined by
\begin{equation}
m_v(s)=
\begin{cases}
\displaystyle\frac{L_v-s_v}{L_v-s_v^{\mathrm{nom}}},
& \text{upper limit},\\[6pt]
\displaystyle\frac{s_v-L_v}{s_v^{\mathrm{nom}}-L_v},
& \text{lower limit},
\end{cases}
\qquad m(s)=\min_{v}m_v(s).
\label{eq:admit_margin}
\end{equation}
The scaling gives $m_v=1$ at the nominal point and $m_v=0$ at the corresponding limit, with negative values indicating a limit violation. T4 rejects a proposal if its projected trajectory crosses a protective limit or if it degrades the minimum margin by more than 0.2 relative to no action:
\begin{equation}
\min_{0\leq t\leq H}m\big(s_{\mathrm{plan}}(t)\big)
<
\min_{0\leq t\leq H}m\big(s_{\mathrm{null}}(t)\big)-0.2,
\qquad H=30\ \mathrm{min}.
\label{eq:admit_t4}
\end{equation}
The relative-margin comparison and the projected-limit check are separate rejection conditions. Both use the same matched simulator assumption as the digital-twin benchmark, but ADMIT T4 is a \emph{constraint-oriented admissibility test}, whereas the digital-twin validator explicitly searches for full recovery or a better mitigation. Their decision rules are therefore not interchangeable even when both use a 30-min model rollout.

A failed gate returns violation codes and suggested remedies for another proposal or a corrected action record. If the record is not admitted after the permitted attempts, no new command is dispatched. In protocol v2, a record that is corrected without changing its commands can be checked again without treating the unchanged commands as a different intervention. Both ADMIT configurations use an LLM planning stage in the reported campaign. The plant-specific ADMIT cartridge includes the system and safety-case graphs, hazards linked to the eleven protective functions, five SOPs, and the twenty benchmark cases; it is generated from the plant definitions to maintain consistency with the benchmark implementation.

\subsection{Rejection handling, fallback, and interpretation}
\label{sec:validation_fallback}

The five configurations share the same operator-level dispatch interface but do not impose the same rejection or fallback policy. Executability-only dispatch applies the first executable proposal. The LLM judge and both ADMIT configurations withhold new supervisory commands after unsuccessful attempts. The digital twin can instead dispatch an eligible counterfactual mitigation that improves on doing nothing, even if none of the candidates meets its full-restoration criterion. These distinctions are retained as part of the \emph{complete validation configurations} being compared rather than obscured by a single acceptance-rate metric.

Rejection feedback is also protocol dependent. In protocol v1, validator failures are returned through the common reprompt path. Protocol v2 additionally provides explicit plain-language feedback identifying the current attempt and previously rejected plans, and records complete LLM exchanges for replay. The experiments therefore compare closed-loop outcomes of agent--validator pipelines with these defined interfaces; they do not isolate every individual contribution of feedback wording, planning, gate logic, and fallback strategy. Section~\ref{sec:protocol} specifies the model and protocol assignments, and Section~\ref{sec:results} reports plant-stop outcomes, economic performance, rejection mechanisms, token use, and wall-clock decision time.

It should be noted that none of the five configurations certifies industrial safety explicitly. The rule-based gates establish compliance with an encoded action policy, not correctness of the policy itself. The projected validators rely on a matched 30-min simulator rollout and do not test robustness to model error or unavailable future information. The LLM judge offers neither deterministic admissibility checks nor physical-state prediction. These limitations are especially pertinent when comparing a method that rejects many proposals with one that permits a larger number of recoveries: the relevant outcome is essentially the resulting plant behaviour and not validation acceptance in isolation.
\section{Benchmark design and evaluation protocol}
\label{sec:benchmark}

The benchmark evaluates the closed-loop consequences of supervisory actions proposed by OperAItor, rather than the linguistic correctness of individual proposals. Each experiment uses the integrated reactor--distillation simulator, regulatory controllers, protective interlocks, and restricted supervisory interface described in Sections~\ref{sec:case}--\ref{sec:framework}. The factors varied across agent experiments are the decision protocol, action-generation model, and complete validation configuration; any planner differences associated with a configuration are treated as part of that configuration. Within a scenario, fault identities, onset and removal times, and the scoring horizon are fixed across configurations. This design allows differences in plant-stop frequency and economic outcome to be interpreted against the same simulated fault conditions, subject to the protocol distinctions discussed below.

\subsection{Fault scenarios}
\label{sec:scenarios}

Twenty scenarios are constructed from the ten benchmark faults in Table~\ref{tab:faults}. They comprise ten single faults (S01--S10), four two-fault cases involving different units or control functions (D01--D04), three coupled cases with overlapping effects (C01--C03), and three triple-fault cases (T01--T03). The triple-fault cases span nominally separate effects (T01), two interacting faults accompanied by a third fault (T02), and three interacting faults (T03). These groupings describe how the cases were designed; the resulting severity is established by the simulated process response rather than assumed from the number of faults alone.

The first fault is injected at $t_0=10$~min. Additional faults, when present, are introduced at the offsets shown in Table~\ref{tab:scenarios}; a missing offset indicates simultaneous onset with the first fault. To determine a common intervention window for each scenario, a \emph{probe} simulation is first conducted with the faults left active and without agent intervention. The probe identifies the first time $t_{\mathrm{trip}}$ at which a plant-stop interlock would activate, if one occurs. The fault-removal times are subsequently fixed by
\begin{equation}
 t_{j}^{\mathrm{lift}}=
 \begin{cases}
 \max\!\left(t_{\mathrm{trip}}+10,\;t_j^{\mathrm{on}}+10\right),
 & \text{if the probe trips},\\
 t_j^{\mathrm{on}}+30,
 & \text{otherwise},
 \end{cases}
 \label{eq:fault_lift}
\end{equation}
where all times are in minutes. The scenario evaluation horizon is
\begin{equation}
 W=\max\!\left(120,\;\max_j t_j^{\mathrm{lift}}+60\right)
 \quad\text{min}.
 \label{eq:scenario_window}
\end{equation}
The removal schedule is therefore determined once from the uncontrolled probe and is not recalculated for each agent or validator. For tripping probes, faults persist beyond the uncontrolled trip time, ensuring that a successful agent must influence the process before that event. An additional period after fault removal permits evaluation of subsequent recovery and inventory restoration. Every 5~min after the final fault is removed, a simulated shift operator requests the reset of any remaining latched interlocks, subject to their normal reset conditions. This common reset mechanism is distinct from the agent's recovery decisions.

\begin{table*}[t]
\centering
\caption{Twenty benchmark fault scenarios. Offsets in parentheses are minutes after the first fault at $t=10$~min. Trip times and interlocks refer to persistent-fault probe simulations without agent intervention, not to the finite-duration benchmark baseline. Parentheses around an interlock indicate a non-trip latch; a dash indicates that no plant-stop interlock was reached in the probe.}
\label{tab:scenarios}
\footnotesize
\renewcommand{\arraystretch}{1.12}
\begin{tabularx}{\textwidth}{@{}lXr l@{}}
\toprule
ID & Fault combination (onset offset) & Probe trip [min] & First trip interlock \\
\midrule
S01 & Reflux pump trip & 16.9 & \id{COL\_L\_LL} \\
S02 & Bottoms pump trip & 48.7 & \id{RB\_L\_HH} \\
S03 & Distillate pump trip & 22.4 & \id{DR\_L\_HH} \\
S04 & Feed pump trip & 81.1 & \id{DR\_L\_LL} \\
S05 & Reactor outlet capacity loss & -- & (\id{R\_L\_HH}) \\
S06 & H$_2$ vent-capacity degradation & 177.8 & \id{RB\_L\_HH} \\
S07 & Reactor heater stuck high & -- & (\id{R\_T\_HH}) \\
S08 & Condenser cooling loss & 23.2 & \id{DR\_L\_LL} \\
S09 & Reflux-drum leak & 46.2 & \id{DR\_L\_LL} \\
S10 & Reactor pressure-sensor low bias & 197.5 & \id{RB\_L\_HH} \\
\midrule
D01 & Pressure-sensor bias + bottoms pump trip (+3) & 49.9 & \id{RB\_L\_HH} \\
D02 & Heater stuck high + distillate pump trip & 21.9 & \id{DR\_L\_HH} \\
D03 & Vent degradation + reflux pump trip (+5) & 21.9 & \id{COL\_L\_LL} \\
D04 & Pressure-sensor bias + drum leak (+2) & 46.7 & \id{DR\_L\_LL} \\
\midrule
C01 & Condenser cooling loss + reflux pump trip (+3) & 20.4 & \id{COL\_L\_LL} \\
C02 & Vent degradation + pressure-sensor bias (+2) & 176.5 & \id{RB\_L\_HH} \\
C03 & Drum leak + distillate pump trip (+4) & 46.8 & \id{DR\_L\_LL} \\
\midrule
T01 & Vent degradation + distillate pump trip (+2) + bottoms pump trip (+4) & 24.9 & \id{DR\_L\_HH} \\
T02 & Condenser loss + pressure-sensor bias (+1) + reflux pump trip (+3) & 20.4 & \id{COL\_L\_LL} \\
T03 & Condenser loss + reflux pump trip (+2) + drum leak (+4) & 19.3 & \id{COL\_L\_LL} \\
\bottomrule
\end{tabularx}
\end{table*}

Eighteen probe scenarios reach a plant-stop interlock. The remaining two, S05 and S07, activate non-trip interlocks but do not produce a plant stop under the persistent-fault probe conditions. In the finite-duration benchmark without agents, 18 of the 20 scenarios also end in a plant stop. The probe and the no-agent baseline serve different purposes: the former establishes fault duration and evaluation horizons, whereas the latter supplies the operational and economic comparator under those fixed durations.

\subsection{Operational and economic evaluation}
\label{sec:economics}

The principal operational outcome is whether the plant remains available throughout the scenario horizon. A plant stop is recorded if a trip-latching protective interlock activates or an episode terminates because of damage, a model-limit violation, or numerical failure. These termination causes are retained separately in the recorded data, because a numerical solver failure is not equivalent to an actual protective intervention, even though both prevent completion of the simulated production episode. Non-trip interlock activation and successful subsequent recovery are not automatically counted as plant stops.

An economic score complements plant-stop counts by measuring the consequences of continued operation, production quality, utility consumption, and inventory changes. The benchmark uses assumed commodity prices, utility charges, and stop penalties for the hypothetical process; these are not measured industrial costs or estimates of commercial profitability. Table~\ref{tab:economic_assumptions} summarises the parameters. All economic calculations use the simulated \emph{true} plant state, rather than potentially biased BPCS measurements. At each 0.5-min observation, the instantaneous operating margin $\dot\Pi$ in \$\,h$^{-1}$ is
\begin{align}
\dot\Pi ={}& \mathbb{1}[x_D\geq0.90]\,
D x_D\frac{M_{\mathrm{ac}}}{1000}p_{\mathrm{ac}} 
\notag\\
& +0.95\,B(1-x_B)\frac{M_{\mathrm{IPA}}}{1000}p_{\mathrm{IPA}}
\notag\\
& +F_{\mathrm{vent}}M_{H_2}p_{H_2}
-F\frac{M_{\mathrm{IPA}}}{1000}p_{\mathrm{IPA}}
\notag\\
& -3.6\,p_{\mathrm{st}}\left(Q_{\mathrm{reb}}+Q_j\right),
\label{eq:economic_margin}
\end{align}
where $D$, $B$, $F$, and $F_{\mathrm{vent}}$ are molar flow rates in kmol\,h$^{-1}$; $x_D$ and $x_B$ are acetone mole fractions; and $M_{\mathrm{ac}}$, $M_{\mathrm{IPA}}$, and $M_{H_2}$ are molecular weights in kg\,kmol$^{-1}$. The prices $p_{\mathrm{ac}}$, $p_{\mathrm{IPA}}$, $p_{H_2}$, and $p_{\mathrm{st}}$ are defined in Table~\ref{tab:economic_assumptions}. Both $Q_{\mathrm{reb}}$ and $Q_j$ are expressed in MW; the factor 3.6 converts MW to GJ\,h$^{-1}$. The factor 0.95 values recovered IPA relative to fresh IPA, while distillate below the specified acetone mole-fraction threshold earns no acetone revenue.

\begin{table}[t]
\centering
\small
\caption{Nominal operating scale and illustrative economic assumptions for the hypothetical benchmark. Monetary coefficients are assumed values in USD}
\label{tab:economic_assumptions}
\begin{tabularx}{\columnwidth}{@{}Xr@{}}
\toprule
Parameter & Value \\
\midrule
Nominal IPA feed rate & 100~kmol/h (6.01~t/h) \\
Nominal acetone product rate & Approx. 4.0~t/h \\
Acetone selling price, $p_{\mathrm{ac}}$ & \$1,300/t \\
Fresh IPA price, $p_{\mathrm{IPA}}$ & \$1,000/t \\
Recovered IPA credit & 95\% of fresh IPA price \\
Hydrogen fuel credit, $p_{H_2}$ & \$1/kg \\
Heating utility, $p_{\mathrm{st}}$ & \$8/GJ \\
On-spec acetone mole fraction & $x_D\geq0.90$ \\
Shutdown charge per plant stop & \$5,000 \\
Restart charge per plant stop & \$15,000 \\
\bottomrule
\end{tabularx}
\end{table}

Let $t_{\mathrm{end}}$ denote the end of the simulated episode, equal to $W$ if operation continues or to the earlier termination time if the plant stops. The scenario-level score is
\begin{align}
\Pi_W={}&
\sum_{k:\,t_k<t_{\mathrm{end}}}
\dot\Pi(t_k)\,\Delta t_{\mathrm{obs}}^{\mathrm{h}}
+V_{\mathrm{inv}}(t_{\mathrm{end}})-V_{\mathrm{inv}}(t_1)
\notag\\
&-\mathbb{1}[\text{plant stop by }W]\,C_{\mathrm{stop}},
\label{eq:scenario_profit}
\end{align}
where $\Delta t_{\mathrm{obs}}^{\mathrm{h}}=0.5/60$~h is the observation interval and $V_{\mathrm{inv}}$ is the value of liquid holdup in the reactor, reboiler, and reflux drum under the benchmark's IPA-equivalent inventory convention. The stop penalty is $C_{\mathrm{stop}}=\$20\,000$, comprising \$5\,000 for shutdown and \$15\,000 for restart. After a stop, no further operating margin is credited within the fixed scenario horizon. The inventory term adjusts the score for the value of material retained in or withdrawn from the modelled process holdups.

For a configuration $c$, the aggregate saving relative to the no-agent baseline is
\begin{equation}
 S_c=\sum_{s=1}^{20}
 \left(\Pi_{W,s}^{(c)}-\Pi_{W,s}^{(\mathrm{no\ agents})}\right).
 \label{eq:aggregate_saving}
\end{equation}
The horizon $W$ is scenario-specific but identical across the configurations compared within that scenario. Aggregated over the twenty scenarios, the nominal fault-free reference has a score of \$34\,202, whereas the no-agent baseline has a score of $-\$364\,878$. Their difference, \$399\,080, is the \emph{nominal-reference performance gap}; it is a comparison point rather than a mathematical upper bound on what a supervisory strategy could earn. The 18 no-agent stops contribute \$360\,000 in fixed shutdown and restart charges, approximately 90\% of this gap. The remaining difference is attributable to the operating-margin and inventory terms. These totals aggregate 20 separate simulated scenarios with horizons defined by Eq.~\eqref{eq:scenario_window}; they are not annual production losses or projected industrial savings. Monetary comparisons should therefore be interpreted alongside the plant-stop outcomes and the assumed trip charges.

\subsection{Decision protocols and agent configurations}
\label{sec:protocol}

Two decision protocols are included. \emph{Protocol v1} initiates a decision after a PCA-confirmed fault and imposes a 5-min cooldown between decisions. As long as the fault remains detected, additional proposals can be generated at subsequent eligible decision points. Protocol v1 uses GPT-5 nano for the action agent and an LLM-based diagnosis stage.

\emph{Protocol v2} uses an event-driven decision schedule. A decision is opened once for a fault episode when an alarm becomes active or the squared prediction error remains above its control limit; a further decision is initiated only by a new priority-1 alarm or a newly failed pump. Within a decision, the action agent receives explicit rejection feedback identifying the attempt number (out of four) and the proposals already rejected. If an ADMIT action record is corrected without changing the commands, the corrected record is rechecked rather than treated as a new intervention. Protocol v2 also retains complete LLM request--response records for replay. Unlike v1, v2 uses the deterministic heuristic diagnoser for every model. This common diagnoser reduces one source of variation when comparing the action-generation models within v2, but the two protocols should not be interpreted as differing only in decision frequency.

Both protocols are evaluated with the five validation configurations defined in Section~\ref{sec:validation} and Table~\ref{tab:validators}: executability-only dispatch, an LLM judge, digital-twin validation, ADMIT T0--T3, and ADMIT T0--T4. Protocol v1 is evaluated with GPT-5 nano \citep{openai2025gpt5}. Protocol v2 is evaluated with GPT-5 nano and three open-weight models served locally using Ollama: Qwen3 8B, Mistral 7B, and Llama 3.2 3B \citep{qwen3,jiang2023mistral,grattafiori2024llama3,ollama}. The action-generation settings described in Section~\ref{sec:action_agent} are applied consistently within the respective configurations. Validation rejection, retry, and fallback behaviour are specified in Section~\ref{sec:validation}, not redefined here. The ADMIT configurations use LLM planning; if other arms use a different planner, the resulting comparison is between complete agent--validator configurations rather than an isolated validator ablation.

The resulting experiment comprises $20\times5=100$ agent runs for protocol v1 and $20\times5\times4=400$ for protocol v2, giving \emph{500 agent runs} in total, in addition to the persistent-fault probes, no-agent baseline, and nominal reference simulations. Thus GPT-5 nano appears in both protocol campaigns, whereas the three open-weight models appear only in v2. Every model--validator--scenario combination is executed once. The principal paired comparisons are between validation configurations \emph{within the same model and protocol}; comparisons between models are made within protocol v2. Cross-protocol differences are descriptive because triggering, diagnosis, and feedback behaviour change together. With only one execution per configuration and scenario, the reported results describe this fixed fault campaign and do not by themselves establish statistical reliability over different random seeds, LLM samples, or disturbance realisations.

In all reported benchmark runs, the simulated process clock is paused while the agent and validator operate, as explained in Section~\ref{sec:agent_architecture}. Inference and validation times are therefore recorded as computational costs rather than inserted as delays into the simulated plant dynamics. The benchmark records plant stops, economic scores, issued and rejected actions, rejection reasons, inference tokens, and decision latency, so that validation effectiveness can be assessed jointly in terms of fault recovery, operational performance, and computational burden.

\section{Results}
\label{sec:results}

The results are organised around three questions: whether the simulated plant and monitoring system provide a credible setting for supervisory fault recovery; how alternative agent--validation configurations affect closed-loop operation; and what operational and computational trade-offs are associated with validation. As established in Section~\ref{sec:protocol}, comparisons are made primarily within a fixed model and decision protocol. Protocols v1 and v2 use different triggering, diagnosis, and rejection-feedback procedures, so their outcomes are not treated as a controlled ablation of any one of these changes.

\subsection{Plant-model verification}
\label{sec:verification}

The process simulator is accompanied by automated physical-consistency and software tests, incorporated into a 178-test suite executed in continuous integration. The audit checks total material accounting during transients involving feed changes, liquid leaks, and reflux interruptions; component and hydrogen balances at the nominal operating point; and consistency of the nominal condenser and reboiler duties. A step-size refinement test using one quarter of the nominal integration step changes the tested temperatures by approximately 0.001~K. These checks support the numerical consistency of the benchmark under the examined operating conditions, rather than establishing fidelity to an industrial plant.

The implemented faults also produce the intended causal responses. For example, a reflux-pump trip reduces reflux, increases top-tray temperature, decreases distillate purity, and can ultimately trigger a low-holdup interlock. Reduced catalyst activity lowers conversion, while condenser cooling loss limits condensation capacity and can deteriorate the column pressure and reflux-drum inventories. Two modelling limitations are relevant to interpretation. First, the reactor uses the irreversible surrogate chemistry described in Section~\ref{sec:case}. Second, the lumped condenser-pressure model does not include a separate vapour-composition inventory; the resulting transient acetone-accounting discrepancy is reported as $C_P P_c\,\Delta y_1$, below 1\% of acetone holdup in the audited cases, and reverses as the top composition returns. Accordingly, subsequent safety and recovery findings refer to this simulator and its defined fault mechanisms.

\subsection{Fault detection and diagnosis}
\label{sec:monitor_results}

\begin{figure*}[t]
\centering
\includegraphics[width=\textwidth]{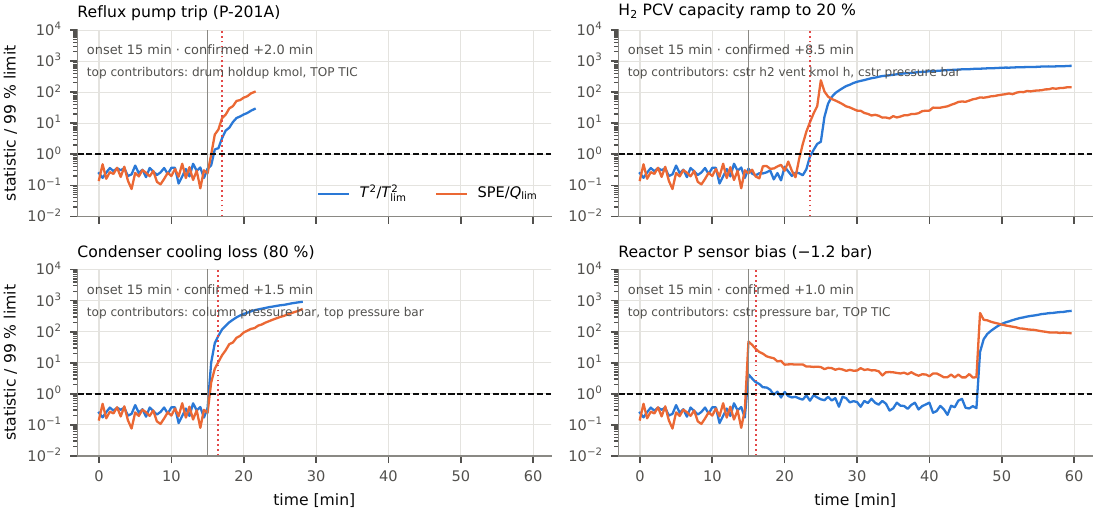}
\caption{PCA monitoring during four representative diagnostic tests with faults introduced at 15~min (vertical grey line). Hotelling's $T^2$ and squared prediction error ($Q$) are normalised by their respective 99th-percentile thresholds (dashed lines). Dotted red markers indicate confirmation after three consecutive threshold violations. Traces terminate when the simulated plant stops.}
\label{fig:pca}
\end{figure*}

Figure~\ref{fig:pca} illustrates the monitoring response in four diagnostic tests, each with a fault injected at 15~min; these tests are separate from the benchmark scenarios whose first fault occurs at 10~min. Neither PCA statistic crossed its detection threshold before fault onset. The reflux-pump trip, condenser cooling loss, and reactor-pressure sensor bias were confirmed within 2.0, 1.5, and 1.0~min, respectively. The progressively degrading hydrogen pressure-control valve (PCV) was confirmed after 8.5~min, once the reduced vent capacity had caused a sufficiently large pressure deviation.

The leading variable contributions were consistent with the affected equipment. Reflux-pump failure increased the contributions of drum holdup and the top-temperature controller output; condenser cooling loss affected the column and overhead pressure measurements; PCV degradation was reflected in reactor pressure and vent flow; and pressure-sensor bias was dominated by the associated pressure measurement. For the sensor fault, the initial signature was principally an increase in squared prediction error ($Q$), indicating a disrupted relationship among measurements, whereas the larger $T^2$ response developed later as the biased pressure reading affected regulatory operation, at approximately 47~min. These examples show that the monitor supplies useful fault evidence, but they do not establish perfect detection or isolation over the entire fault library.

At the first PCA confirmation in each of the 20 benchmark scenarios, the heuristic diagnoser ranked at least one true injected fault first in 14 scenarios. The corresponding numbers for GPT-5 nano and GPT-5 mini diagnostic evaluations were 15 and 18. All three approaches included at least one true fault among their top three hypotheses in all 20 scenarios. Most top-ranked errors arose from faults with similar early symptoms, including vent restrictions and pressure-measurement bias, or from upstream disturbances whose initial column-side effects resembled condenser problems. GPT-5 mini is included here as a diagnosis-only comparison; it is not one of the four action-generation models in the principal control campaign. The downstream experiments consequently test action generation and validation under plausible but imperfect diagnostic evidence.

\subsection{Closed-loop fault-recovery performance}
\label{sec:campaign_results}

\begin{figure*}[t]
\centering
\includegraphics[width=\textwidth]{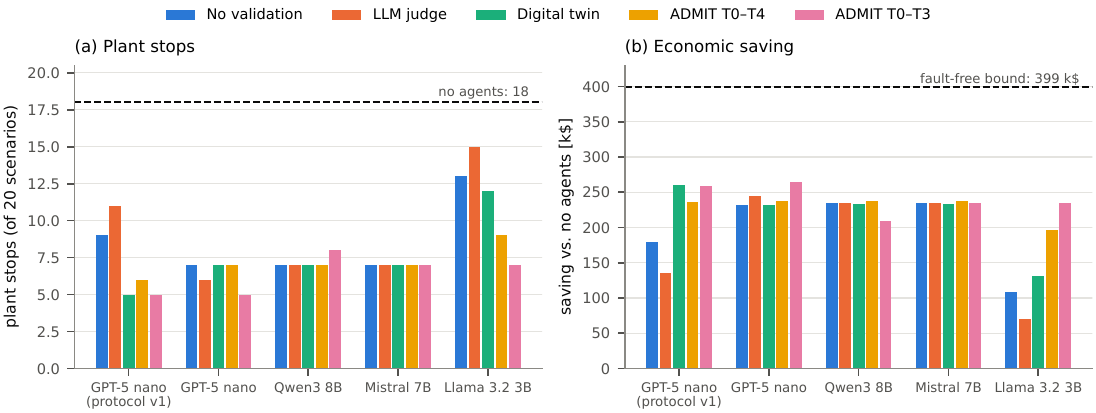}
\caption{Campaign outcomes across action-generation models and validation configurations: (a) plant stops out of 20 scenarios and (b) aggregate savings relative to the no-agent baseline. GPT-5 nano is shown separately for protocols v1 and v2 because the protocols differ in diagnosis, triggering, and feedback. The savings reference line represents the \$399,080 nominal-reference performance gap, not a theoretical maximum.}
\label{fig:bars}
\end{figure*}

The principal campaign comprises 500 episodes with agents: 100 with GPT-5 nano under protocol v1, and 400 under protocol v2 across four action-generation models. Each model is evaluated with five validation configurations and 20 scenarios. The no-agent baseline stops in 18 scenarios. Table~\ref{tab:campaign} reports the number of plant stops, the aggregate saving relative to the no-agent baseline, and the mean decision wall time. Figure~\ref{fig:bars} provides the corresponding graphical comparison. The nominal-reference gap of \$399,080 (Section~\ref{sec:economics}) is used as a reference level in the savings plot; it is not an upper bound on achievable performance.

\begin{table*}[t]
\centering
\caption{Results over 20 scenarios for each model, protocol, and validation configuration. Savings are relative to the no-agent baseline (k\$), and times are mean wall-clock seconds per decision. Each row represents 20 episodes, with one execution per scenario.}
\label{tab:campaign}
\small
\setlength{\tabcolsep}{4pt}
\renewcommand{\arraystretch}{1.08}
\begin{tabular}{@{}llrrr@{}}
\toprule
Model and protocol & Validation & Plant stops & Saving [k\$] & Time [s/decision]\\
\midrule
\multicolumn{5}{@{}l}{\textit{Protocol v1: repeated decisions; LLM diagnosis}}\\
GPT-5 nano & None & 9 & 179 & 21\\
 & LLM judge & 11 & 135 & 29\\
 & Digital twin & 5 & 260 & 167\\
 & ADMIT T0--T4 & 6 & 237 & 143\\
 & ADMIT T0--T3 & 5 & 259 & 42\\
\midrule
\multicolumn{5}{@{}l}{\textit{Protocol v2: event-driven decisions; heuristic diagnosis}}\\
GPT-5 nano & None & 7 & 233 & 8\\
 & LLM judge & 6 & 245 & 20\\
 & Digital twin & 7 & 232 & 118\\
 & ADMIT T0--T4 & 7 & 237 & 99\\
 & ADMIT T0--T3 & 5 & 265 & 28\\
\addlinespace
Qwen3 8B & None & 7 & 235 & 66\\
 & LLM judge & 7 & 235 & 96\\
 & Digital twin & 7 & 234 & 336\\
 & ADMIT T0--T4 & 7 & 238 & 398\\
 & ADMIT T0--T3 & 8 & 209 & 219\\
\addlinespace
Mistral 7B & None & 7 & 235 & 76\\
 & LLM judge & 7 & 235 & 66\\
 & Digital twin & 7 & 234 & 405\\
 & ADMIT T0--T4 & 7 & 238 & 361\\
 & ADMIT T0--T3 & 7 & 235 & 196\\
\addlinespace
Llama 3.2 3B & None & 13 & 109 & 138\\
 & LLM judge & 15 & 70 & 164\\
 & Digital twin & 12 & 131 & 109\\
 & ADMIT T0--T4 & 9 & 197 & 191\\
 & ADMIT T0--T3 & 7 & 235 & 90\\
\bottomrule
\end{tabular}
\end{table*}

\paragraph{Repeated supervisory decisions (protocol v1).}
Under protocol v1, GPT-5 nano could make a new decision every 5~min while a fault remained detected. The executability-only agent stopped the plant in nine scenarios, whereas the LLM-judge configuration stopped it in eleven. The digital twin and ADMIT T0--T3 configurations each stopped the plant in five scenarios; ADMIT T0--T4 stopped it in six. Relative to no agents, the corresponding savings were \$179k (no validation), \$135k (judge), \$260k (digital twin), \$259k (T0--T3), and \$237k (T0--T4). Thus, in this fixed repeated-decision campaign, simulation-based validation and admissibility checking improved the operational outcome over unvalidated dispatch. The LLM judge did not: it permitted some proposals associated with subsequent trips and rejected other proposals that could have supported recovery. These observations describe complete pipeline outcomes, not independently measured false-acceptance or false-rejection probabilities.

\paragraph{Event-driven decisions with larger models (protocol v2).}
GPT-5 nano, Qwen3 8B, and Mistral 7B produced relatively similar outcomes across most validation configurations, with five to eight plant stops and \$209--265k in savings. The best observed v2 result was GPT-5 nano with ADMIT T0--T3, which stopped the plant in five scenarios and saved \$265k. For Qwen3 8B, however, T0--T3 resulted in eight stops and \$209k compared with seven stops and \$235k without validation; Mistral 7B retained seven stops in every configuration. Consequently, the results do not establish that one validator dominates across capable models. Differences in the intervention selected and the time consumed by validation can matter even when aggregate stop counts are similar.

\paragraph{Sensitivity to the action-generation model.}
The clearest variation among the v2 models occurred with Llama 3.2 3B. It stopped the plant in 13 scenarios without validation, 15 with the LLM judge, and 12 with the digital twin. ADMIT T0--T4 reduced the number to nine, while T0--T3 reduced it to seven and increased savings to \$235k, comparable to the observed totals for the 8B and 7B models. In the four single-pump-trip cases (S01--S04), the T0--T3 configuration started a standby pump within 2.5~min of the fault, whereas the unvalidated 3B configuration issued no intervention in those cases. One plausible mechanism is that the structured ADMIT action record and associated rejection feedback guide the smaller model towards an actionable command. However, the experiments do not isolate this prompting effect from the acceptance gates, planner, or reprompting procedure. The result should therefore be interpreted as an improvement of the complete ADMIT-enabled configuration, not proof that deterministic checking alone improves small-model reasoning.

\subsection{Scenario-level recovery and failure modes}
\label{sec:scenario_results}

\begin{figure*}[t]
\centering
\includegraphics[width=\textwidth]{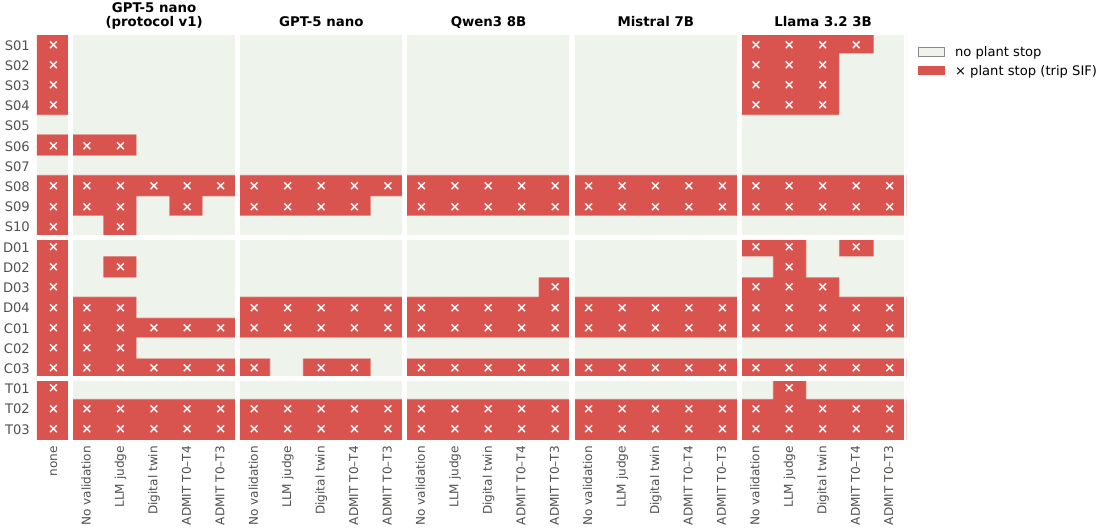}
\caption{Plant-stop outcome for each scenario and agent--validation configuration. The no-agent reference is shown separately. A cross denotes an episode ending in a plant stop; an unmarked cell denotes operation without a recorded stop over the benchmark horizon.}
\label{fig:grid}
\end{figure*}

The aggregate results conceal marked differences in scenario difficulty. Figure~\ref{fig:grid} shows the outcome of each model--validator combination alongside the no-agent baseline. Individual duty-pump trips (S01--S04) are generally recoverable when an agent identifies the failed train and uses the KG-linked \id{START\_SPARE} command. The same mechanism can mitigate pump trips in multi-fault cases such as D01--D03 and T01, although simultaneous faults create additional constraints. Slowly developing reactor faults, including hydrogen PCV degradation (S06), pressure-sensor bias (S10), and their combination (C02), require an intervention that maintains acceptable operating conditions across the coupled reactor and column after the initial disturbance.

\begin{figure*}[t]
\centering
\includegraphics[width=\textwidth]{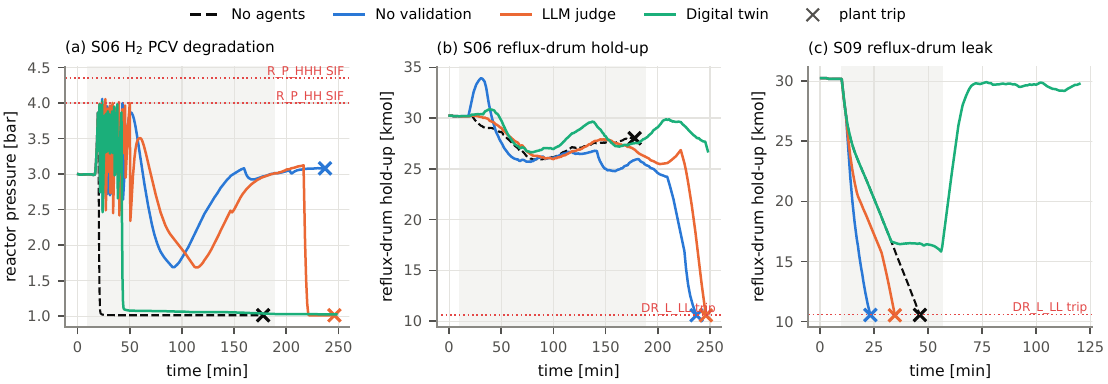}
\caption{Representative closed-loop trajectories for GPT-5 nano under protocol v1: (a,b) reactor pressure and reflux-drum holdup during S06 (hydrogen-vent degradation), and (c) reflux-drum holdup during S09 (drum leak). Shaded regions indicate active faults, dotted red lines denote protective limits, and crosses indicate plant stops. The S06 persistent-fault probe stops at approximately 178~min.}
\label{fig:traj}
\end{figure*}

Figure~\ref{fig:traj} provides illustrative protocol-v1 trajectories with GPT-5 nano. In S06, repeated throughput reductions and pressure-setpoint changes by the unvalidated and LLM-judge agents destabilised the coupled operating balance. Their episodes ended in reflux-drum low-low trips at approximately 237 and 246~min, respectively, after the injected fault had been removed. By contrast, a digital-twin-selected mitigation maintained operation over the evaluated horizon. This case illustrates that avoiding the immediate fault signature is insufficient if the intervention depletes an inventory elsewhere in the process. In S09, the unvalidated agent lowered the reflux-drum level setpoint while the drum was leaking, thereby increasing net withdrawal and accelerating the trip to approximately 23~min, compared with approximately 46~min without an agent. The digital-twin configuration instead selected a throughput reduction and higher top-temperature setpoint, reducing reflux demand and maintaining drum inventory until the scheduled leak removal. These examples show why an intervention's effect must be assessed across connected units rather than solely against the measurement that triggered the decision.

Seven scenarios were particularly difficult across the campaign: the four involving condenser cooling loss (S08, C01, T02, and T03) and three strongly affected by reflux-drum leakage (S09, D04, and C03). Most configurations stopped in these cases, although GPT-5 nano with ADMIT T0--T3 under protocol v2 avoided a stop in S09. These observations identify a demanding subset of the benchmark; they do not, on their own, establish that every possible supervisory strategy is incapable of recovering all seven scenarios. The more restricted single-action feasibility question is examined next.

\subsection{Offline admissibility and action-authority limits}
\label{sec:admit_offline}

The offline ADMIT evaluation uses one case per benchmark scenario at the time of PCA confirmation, together with an associated hazard definition and a catalogue of candidate supervisory interventions. For cases whose persistent-fault probe trips the plant, the observed trip interlock provides the principal hazard. An oracle policy selects the admissible candidate with the greatest projected safety margin from standby-pump starts, throughput changes, blowdown, emergency cooling, and top- or bottom-temperature setpoint changes. If no candidate satisfies the safety envelope, it selects escalation rather than an unsupported plant intervention.

For the four condenser-cooling-loss cases (S08, C01, T02, and T03), no evaluated \emph{single} permitted action maintained the projected trajectory within the prescribed envelope. The oracle therefore escalated. This outcome is consistent with the observed loss of condenser duty and deterioration of pressure or drum-inventory conditions, however it does not rule out multi-action recovery policies, additional equipment, or a different protection design. It also distinguishes a desirable controlled shutdown or escalation from successful continued production.

In the offline harness, the oracle records were admitted in all 20 cases, with a reported mean aggregate score of 0.998; an escalation can be admissible even when it does not recover production. A test agent generating predefined inadmissible proposals was blocked in 12 of 20 cases: four at T0 for attempting to bypass a protective function, six at T1 for requesting blowdown or emergency cooling without the required evidence, and two at T3 for missing prescribed procedural steps on setpoint changes. These checks illustrate the gate mechanism's ability to reject specified authority, evidence, and procedure violations in the offline test. They are not a statistical estimate of its safety performance on arbitrary proposals.

\subsection{Rejection behaviour and computational cost}
\label{sec:rejections_cost}

\begin{figure*}[htbp]
\centering
\includegraphics[width=0.72\linewidth]{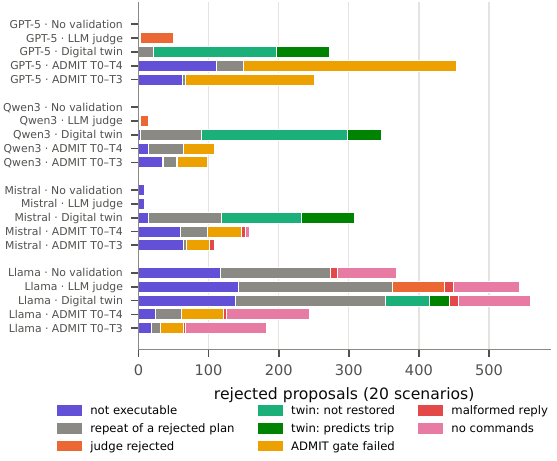}
\caption{Rejection causes under protocol v2, grouped by model and validation configuration. Each configuration is evaluated on 20 scenarios; the total number of proposals can differ because rejected actions may be revised and resubmitted.}
\label{fig:rej}
\end{figure*}

Figure~\ref{fig:rej} groups proposal rejections by cause under protocol v2. The digital-twin validator commonly rejected proposed full-recovery actions when its 30-min projection failed to restore the active targets or predicted a plant stop. ADMIT rejections were more often associated with incomplete or inconsistent action records, including missing or incorrectly ordered procedure steps (T3) and citations to evidence absent from the trusted ledger (T1). In protocol v1 with GPT-5 nano, only 3\% of digital-twin decisions accepted the first proposal under the default full-recovery criterion, indicating that revisions and the fallback mitigation mechanism were important parts of that configuration. The different rejection distributions reflect different acceptance questions: predicted process consequence for the twin, and documented admissibility conditions for the gates. Rejection counts alone therefore cannot rank validation effectiveness.

\begin{figure*}[t]
\centering
\includegraphics[width=\textwidth]{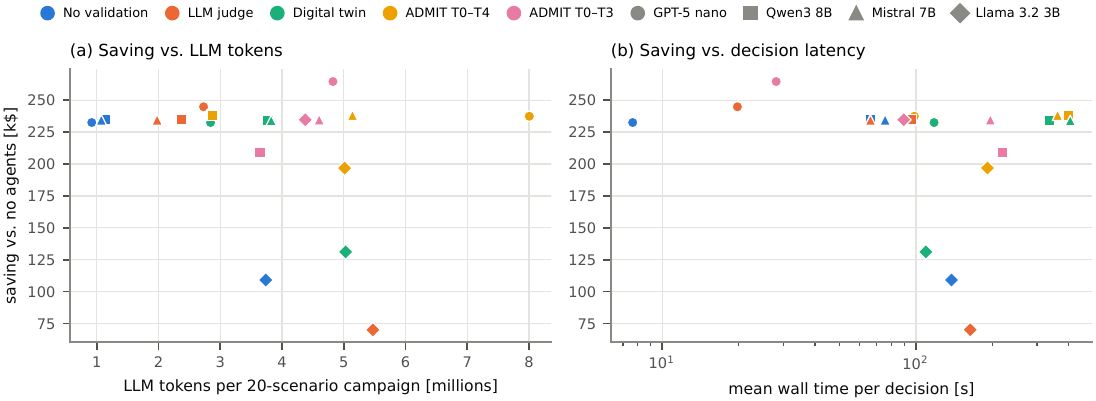}
\caption{Protocol-v2 operating-point comparison: aggregate saving versus (a) total LLM tokens per 20-scenario configuration and (b) mean wall-clock time per decision. Markers distinguish action-generation models and validation configurations. The points are descriptive campaign outcomes rather than estimates of expected performance across random seeds.}
\label{fig:cost}
\end{figure*}

The performance benefits of validation must be considered alongside its computational requirements. Figure~\ref{fig:cost} relates savings to token use and wall-clock decision latency for protocol v2. With GPT-5 nano, the 20-scenario totals were approximately 0.9 million tokens for no validation, 2.7 million for the LLM judge, 2.8 million for the digital twin, 8.0 million for ADMIT T0--T4, and 4.8 million for ADMIT T0--T3. Mean decision times were approximately 7.7, 20, 118, 99, and 28~s, respectively. The reported API charge for the GPT-5 nano no-validation configuration was approximately \$0.08, and the full 100-run GPT-5 nano v2 campaign cost \$1.58 in metered API usage at the time of the experiments. These charges exclude local simulation, hardware, electricity, and development costs.

For the GPT-5 nano digital-twin configuration, forward simulation accounted for approximately 117~min of the campaign's 147~min total decision time. Local Qwen3, Mistral, and Llama inference incurred no metered model-API charge, but individual LLM calls took approximately 22--115~s on the tested GPU, excluding the unreported economic cost of local compute. ADMIT T0--T3 achieved the highest v2 saving for GPT-5 nano and Llama 3.2 3B at lower mean latency than their digital-twin or T0--T4 configurations. That operating point was not universal: for Qwen3 8B, T0--T3 reduced savings and increased the stop count, while the Mistral configurations had nearly identical economic outcomes despite substantially different latencies. Thus, the preferred validation configuration depends on the action-generation model and the operational objective; the available data do not identify a uniformly dominant validator.

\subsection{Additional prompt-context ablation and limitations}
\label{sec:prompt_ablation}

A supplementary prompt-context ablation examined six selected protocol-v1 failure scenarios for GPT-5 nano, using repeated trials and also testing GPT-5 mini. Adding loop information, an operating manual, or further process-dynamics descriptions did not consistently improve the outcomes in this selected subset. The reported success counts ranged from zero to four non-stopping runs out of twelve in the KG-supported conditions, compared with zero out of eight in the conditions without the KG; these denominators differ and should not be interpreted as a matched statistical comparison. In several trials, both models continued to propose increased reflux while the drum was draining, despite receiving relevant loop and inventory information. This result suggests that more context or a larger model does not necessarily resolve incorrect intervention selection in these examples, motivating the explicit evaluation of action consequences.

Several limitations constrain the conclusions. First, only one run was conducted for each model--validation--scenario combination in the main campaign, so the reported differences are descriptive and do not establish their variability across repeated LLM samples, initialisations, or fault realisations. Second, validator comparisons can include changes in planning and action-record requirements; where the planner differs, they compare complete agent--validator configurations rather than isolating the validator alone. Third, protocol v1 and v2 differ simultaneously in diagnosis, triggering, and rejection feedback, preventing a causal attribution of their aggregate differences to any one change. Fourth, the digital twin is a matched clone of the simulated plant and is therefore an oracle relative to realistic model mismatch; all decisions are also made with the simulated process clock paused. Finally, the process chemistry, protective thresholds, and economic penalties are benchmark assumptions rather than validated industrial design values.

Within these boundaries, the results support three conclusions. Validation can materially change the outcome of repeated supervisory interventions, particularly when an agent proposes actions that worsen inventories elsewhere in the process. Structured admissibility requirements can improve the performance of a smaller action-generation model in the tested configurations, but admission is not a guarantee for beneficial process consequences. Most importantly, the number of actions accepted or rejected is not itself an operational objective: a useful validation system must be evaluated in terms of avoided stops, preserved recovery opportunities, economic performance, and decision cost. These results motivate further tests under uncertain digital twins, asynchronous execution, and repeated stochastic trials rather than claims of safety certification from the present campaign.
\section{Discussion}
\label{sec:discussion}

\paragraph{Validation outcomes.}
The effect of validation depends on the action-generation model and the
complete decision pipeline. Under protocol v1, repeated intervention by
GPT-5 nano led to fewer plant stops with the digital-twin and ADMIT
configurations than with executability-only dispatch. Under protocol v2,
the results were less consistent: several configurations gave similar
outcomes for GPT-5 nano, Qwen3 8B, and Mistral 7B, while ADMIT T0--T3
increased the stop count for Qwen3 8B. The most pronounced improvement
was observed for Llama 3.2 3B with ADMIT T0--T3. These comparisons also
reflect configuration-specific planning, action records, rejection
feedback, and fallback behaviour; they do not isolate the effect of a
validation algorithm. The LLM judge worsened outcomes for GPT-5 nano
under protocol v1 and Llama 3.2 3B under protocol v2, but its effect was
small or absent in several other configurations.

\paragraph{Digital twin and admissibility gates.}
The digital twin projects candidate actions over 30~min using a matched
copy of the simulated plant. Its results depend on having an accurate
current state and dynamics; plant--model mismatch and state-estimation
errors were not evaluated. In contrast, ADMIT T0--T3 checks authority,
evidence, hazards, and prescribed operating procedures without requiring
a dynamic model. ADMIT T4 supplements these checks with a consequence
projection. The methods therefore test different properties of an
intervention: compliance with encoded operating requirements and
predicted process behaviour. Neither establishes safety beyond the
assumptions of its validation mechanism.

\paragraph{Operational and recovery limits.}
Seven scenarios involving condenser cooling loss or reflux-drum leakage
were particularly difficult across the campaign, although one drum-leak
scenario was recovered by a tested configuration. The offline analysis
found that none of the evaluated single supervisory actions could keep
the four condenser-cooling-loss scenarios within the projected safety
envelope. This does not establish infeasibility for sequences or
combinations of actions, additional equipment, or alternative plant
designs. The recovery limits should therefore be interpreted relative
to the tested action catalogue and its projected horizon.

\paragraph{Study limitations.}
The reactor--distillation process is a hypothetical operating benchmark
with physically based mixture properties and modified liquid-phase
reaction kinetics, rather than a validated acetone manufacturing plant.
The stage-equilibrium and hydraulic models also omit several industrial
phenomena, including tray flooding, weeping, and detailed pump curves.
The economic assumptions, notably the fixed \$20,000 plant-stop charge,
are illustrative and affect the magnitude of the monetary results.
Each main-campaign configuration was executed once per scenario;
therefore, the reported differences do not provide statistical estimates
of run-to-run variability. The process clock was paused during agent
and validator execution. Finally, the reported results use simulator
version 1.0; version 1.1 changes the reboiler steam-valve range and
certain setpoint limits while retaining the nominal duty, so its
benchmark scores may differ.

\paragraph{Benchmark extensibility.}
The released framework allows alternative plants, faults, knowledge
graphs, prompts, validators, and decision pipelines to be evaluated
using the same measurement and scoring interface. Human-operator and
agent results can also be compared on a common scenario set, subject to
matching information access and timing assumptions.

\section{Conclusions}
\label{sec:conclusions}

OperAItor provides an open benchmark and framework for evaluating
LLM-generated supervisory actions in a dynamic reactor--distillation
process equipped with regulatory control, alarms, and simulated
independent protective functions. Five action-validation configurations
were compared using four language models and 20 single- and
multiple-fault scenarios. The no-agent baseline ended in 18 plant
stops. Across GPT-5 nano, Qwen3 8B, and Mistral 7B, the evaluated
validated configurations resulted in 5--8 stops and recovered
\$209,000--265,000 of the \$399,080 nominal-reference performance gap
under illustrative economic assumptions. The largest effect of
admissibility checking was observed for Llama 3.2 3B, for which ADMIT
T0--T3 reduced stops from 13 to 7. No validator dominated all model
and protocol combinations, and LLM-based review worsened outcomes in
some configurations. Four condenser-cooling-loss scenarios had no
feasible recovery among the evaluated single supervisory actions in
the offline admissibility analysis. Future work will evaluate model
mismatch and state estimation for consequence projection, repeated
stochastic runs, agent-decision latency during plant evolution, and
additional plants and supervisory action policies.

\section*{Code and data availability}
The simulator, agents, knowledge graph, ADMIT cartridge, benchmark harness and all campaign records (every run, LLM request and ADMIT trace) are available under the Apache-2.0 licence at \url{https://github.com/AISL-at-Imperial-College-London/operAItor} (AISL, Imperial College London, 2026a). The project website (AISL, Imperial College London, 2026b), \url{https://aisl-at-imperial-college-london.github.io/operAItor/}, hosts the live console, the interactive campaign dashboard and installers for Windows, macOS and Linux. 

\section*{CRediT authorship contribution statement}
\textbf{Javal Vyas:} Conceptualization, Methodology, Software, Investigation, Formal analysis, Visualization, Writing -- original draft. \textbf{Mehmet Mercang{\"o}z:} Supervision, Funding acquisition,
Writing -- review and editing.

\section*{Declaration of competing interest}
The authors declare that they have no known competing financial
interests or personal relationships that could have appeared to
influence the work reported in this paper.

\section*{Declaration of generative AI and AI-assisted technologies in the manuscript preparation process}
During the preparation of this work, the authors used OpenAI's ChatGPT
to assist with language refinement, manuscript organization, and
drafting support. After using this tool, the authors reviewed and edited
the content as needed and take full responsibility for the content of
the published article.

\section*{Acknowledgements}
The authors acknowledge support from Imperial College London.

\bibliographystyle{elsarticle-harv}
\bibliography{reference}

\end{document}